%% file: The_shareability_of_steering_in_two-producible_states.tex
\documentclass[
pra,aps,
reprint,
a4paper,
superscriptaddress,
longbibliography,
floatfix
]{revtex4-1}
\usepackage{times}
\usepackage{graphicx}
\usepackage{epsfig}
\usepackage{amsfonts}
\usepackage{amsmath}
\usepackage{amssymb}
\usepackage{color}
\usepackage{multirow}
\usepackage[colorlinks=true,linkcolor=blue,citecolor=magenta,urlcolor=blue]{hyperref}
\usepackage{soul}
\usepackage[dvipsnames]{xcolor}
\usepackage{hyperref} 
\usepackage{physics}
\usepackage{dsfont}
\usepackage{tabularx}
\usepackage{makecell}
\usepackage{multirow,array,hhline}
\usepackage{tikz}

\newcolumntype{M}{>{$\vcenter\bgroup\hbox\bgroup}c<{\egroup\egroup$}}
\setstcolor{orange}

\newtheorem{theorem}{Theorem}
\newtheorem{corollary}{Corollary}

\begin{document}

\title{The shareability of steering in two-producible states}

\author{Qiu-Cheng Song}
\author{Travis J. Baker}
\author{Howard M. Wiseman}
\email{h.wiseman@griffith.edu.au}
\affiliation{Centre for Quantum Computation and Communication Technology (Australian Research Council), \\  Centre for Quantum Dynamics, Griffith University, Yuggera Country, Brisbane 4111, Australia}

\begin{abstract}
Quantum steering is the phenomenon whereby one party (Alice) proves entanglement by ``steering'' the system of another party (Bob) into distinct ensembles of states, by performing different measurements on her subsystem.  
Here, we investigate steering in a network scenario involving $n$ parties, who each perform local measurements on part of a global quantum state, that is produced using only two-party entangled states, and mixing with ancillary separable states.
We introduce three scenarios which can be straightforwardly implemented in standard quantum optics architecture, which we call random $\frac{n}{2}$-pair entanglement, random pair entanglement and semi-random pair entanglement.  
We study steerability of the states across two-party marginals which arise in the three scenarios, and derive analytically the necessary and sufficient steering criteria for different sets of measurement settings. 
Strikingly, using the semi-random pair entanglement construction, one party can steer every one of the $n-1$ other parties, for arbitrarily large $n$, using only two measurements. 
Finally, exploiting symmetry, we study various small network configurations (three or four parties) in the three scenarios, under different measurements and produced by different two-party entangled states. 
\end{abstract}
\maketitle

\section{Introduction}

The Einstein, Podolsky, and Rosen (EPR) paradox~\cite{EPR35} involved a bipartite quantum system prepared in an entangled state shared between two parties, whom we will refer to as Alice and Bob. 
Since, by performing different measurements on her sub-system, Alice can affect the possible conditional  states of Bob's system, 
EPR believed that the ambiguity that arises in describing Bob's state was paradoxical.
In his reaction to this paper, Schrödinger~\cite{Sch35} coined the term \emph{steering} to describe this phenomenon. 
EPR-steering was subsequently formalized as a quantum information task, denoting the capability to generate a set of ensembles remotely that cannot be simulated semi-classically through a local hidden state (LHS) model~\cite{Wis07, Jon07}. 
That is, it rules out the possibility that each ensemble held by Bob is locally described by quantum systems, of which Alice has knowledge which she uses to cleverly announce outcomes of her apparent measurements to simulate steering Bob.
EPR-steering, as it has been called~\cite{Caval09}, is known to be a type of nonlocality that is stronger than quantum entanglement, but weaker than Bell nonlocality. 
A notable characteristic that sets EPR-steering apart from other types of nonlocality is its inherent asymmetry.

Detection of EPR-steering is fundamental problem for understanding the steerability of quantum states. 
While the simplest two-qubit entangled states have been extensively studied (see, e.g., Refs.~\cite{Jev14,Mil14,Bak18,Bak20,Yu18,Yu18a,Bow16,Ngu16,Rei09,Jev15,Ngu16a,Quan16,McCl17,Shuming16,Chen17,chenjl13,song22,Uol20}), detecting steering can still be a challenging problem. 
Similar to Bell-CHSH inequalities~\cite{bell1964,chsh69}, EPR-steering can be detected by examining the violation of steering inequalities~ (see e.g.~\cite{Caval09,Caval15,Caval13,Gird16,Quan17}).
An alternative method is to investigate whether a given assemblage~\cite{Pus13}, representing the steered ensembles for Bob, can be reproduced using local hidden states. When the number of measurements and outputs is finite, this problem can be solved using semi-definite programming (SDP) techniques~\cite{Hir16,Fil18,Cav16a}. However, it's important to note that the size of the SDP grows exponentially with the number of measurements, which can pose a challenge; see~\cite{Cav16} for a review.
However, for an infinite number of projective measurements, by utilizing the principles of convex geometry, EPR-steering can be converted into an inclusion problem. Through the application of a linear program, it has been demonstrated that this approach is capable of accurately determining the steerability of nearly all two-qubit states with a high level of precision~\cite{Ngu19}.

In multipartite systems, there exists a number of criteria to detect different types of steering~\cite{Cava11,Heq11,he13,Cavalcanti15,Jones21,Costa18,Xiang22}. Monogamy and shareability are two complementary properties of multipartite steering that can arise. 
Similar to the monogamy in entanglement~\cite{Coffman00} 
and Bell nonlocality~\cite{Kurzy11,Cheng17,Zhu19},
the monogamy of EPR-steering refers to the situation where two parties cannot simultaneously steer the state of the third party~\cite{Reid13}. 
The difference is the inherent directionality of EPR-steering monogamy. More monogamy relations for steering were proposed in Gaussian regime~\cite{JiNha15,Adesso16,Lami16}. However, monogamy of steerability can be broken by increasing the number of measurement settings~\cite{Paul20} or performing non-Gaussian measurements~\cite{JiNha16}.  Unlike the monogamy of EPR-steering, the shareability of EPR-steering in reduced subsystems allows the state of one party to be steered by two or more parties, which shows more configurations of multipartite EPR-steering. 
Recently, shareability of multipartite EPR-steering was demonstrated in an optical experiment~\cite{Hao22}.

In this paper, we study EPR-steeribility of $k$-producible $n$-partite entangled state.  A $n$-partite pure state $|\varphi\rangle$ is termed $k$-producible if it can be written as $|\varphi\rangle=|\phi_1\rangle\otimes|\phi_2\rangle\otimes\cdots\otimes|\phi_f\rangle$, where the states $|\phi_i\rangle$ are states of maximally $k$ parties~\cite{guhne05}. A mixed state is $k$-producible if it can be  written as a mixture of $k$-producible pure states. Here, we try to use minimum resources in terms of (a) minimizing $k$ in the $k$-producibility of our $n$-partite entangled state; (b) minimizing the local dimension of entangled state; (c) minimizing the number of entangled state[s] we have to produce. Remarkably we find steering of arbitrarily many parties is possible with the minimum of resources: the $n$-partite state comprises only qubits locally, is $2$-producible, and can be produced from a \emph{single} entangled pair of qubits.

We come to this conclusion by studying the shareability of steering in three different scenarios of 2-producible multipartite entangled qubit states, which we call random $\frac{n}{2}$-pair entanglement (R$\frac{n}{2}$PE), random pair entanglement (RPE), and semi-random pair entanglement (SRPE). As well as considering steerability under all projective measurements, we consider more limited measurement strategies, for which we find the necessary and sufficient steering criterion analytically for the relevant class of reduced two-qubit states, which are two-qubit X-states. 
Most strikingly, in the SRPE scenario, where the $n$-qubit state can be produced from a single entangled pair of qubits plus $n-2$ product states, one party (Alice) can simultaneously steer all $n-1$ Bobs, for arbitrary $n$, using only two measurements.  
Finally, we study the properties of small networks in the three scenarios.

This paper is organized as follows.
In Sec.~\ref{sec:EPR-steering} we provide an overview of concepts related to EPR-steering and describe a numerical approach for determining which two-qubit states are capable of exhibiting EPR-steering and which are not. 
In Sec.~\ref{sec:The steerability of X-states} we derive analytical criteria for EPR-steering in two-qubit states, subject to certain restriction and varying measurement strategies. These criteria are both necessary and sufficient for the corresponding conditions.
In Sec.~\ref{sec:Multiparty Steering} We present three distinct scenarios and analyze their steerability under various measurement settings.
In Sec.~\ref{sec:networks} we study three- and four-parties network properties in the three scenarios.
In Sec.~\ref{sec:Conclusion} we summarize the main results and give possible future research directions.

\section{Preliminaries}
\label{sec:EPR-steering}

Consider a bipartite quantum state $\rho_{\mathrm{AB}}$ shared between two parties, Alice and Bob. We define Alice's \emph{measurement strategy} ${\cal M}$ as the set of measurements she can perform on her local subsystem. In general this is described by positive operator valued measures (POVMs): 
${\cal M} = \{M_{r \mid s}\}_{r,s}$. These POVMs must satisfy the following conditions: $M_{r \mid s} \geqslant 0$ and $\sum_{r} M_{r \mid s}=\openone$ for all measurement settings $s$ and measurement results $r$. When Alice can perform these measurements on her part of the shared state, Bob's subsystem can be transformed into a collection of states $\{\rho_{r|s}\}_{r,s}$ with probabilities $p_{r|s}$. This collection of states can be represented by an \emph{assemblage} $\{\sigma_{r|s}\}_{r,s}$, where $\sigma_{r|s} := p_{r|s}\rho_{r|s}$. The concept of an assemblage, introduced by Pusey~\cite{Pus13}, represents all the information relevant to EPR-steering in this scenario. 

According to quantum theory, the members of the assemblage can be obtained by
\begin{equation}
\sigma_{r \mid s}=\operatorname{Tr}_{\mathrm{A}}\left[\left(M_{r \mid s} \otimes \openone\right) \rho_{\mathrm{AB}}\right]
\end{equation}
with probability 
\begin{equation}
p_{r|s}=\operatorname{Tr}\left[\left(M_{r \mid s} \otimes \openone\right) \rho_{\mathrm{AB}}\right]. 
\end{equation}
The assemblage satisfies
\begin{align}
\sum_r\sigma_{r | s}=\sum_r\sigma_{r| s^\prime}=\rho_\mathrm{B}\quad \forall s, s^\prime
\end{align}
and 
\begin{align}
\operatorname{Tr}\sum_r\sigma_{r|s}=1\quad \forall s,  
\end{align}
where $\rho_{\mathrm{B}}=\operatorname{Tr}_\mathrm{A}[\rho_\mathrm{AB}]$ is Bob's reduced density matrix.
The assemblage $\{\sigma_{r|s}\}_{r,s}$ 
 is \emph{non-steerable} if and only if there exists a LHS model~\cite{Wis07} that can reproduce the assemblage, meaning that all the member of the assemblage admits the decomposition
\begin{equation}\label{asslhs}
\sigma_{r | s}=\sigma_{r | s}^{\text{LHS}}=\int \mathrm{d} \lambda \mu(\lambda) p(r | s,\lambda) \rho_{\lambda},
\end{equation}
where $\lambda$ is a classical variable with the probability distribution $\mu(\lambda)$,  $\rho_{\lambda}$ is an LHS indexed by $\lambda$, and  $p(r|s, \lambda)$ is the conditional probability distribution. 

A common scenario to consider, especially for qubits, is when Alice's measurements are projective. That is, they can be described a measurements of  observables $O_k$ (Hermitian operators), in which case we denote her measurement scenario more simply as ${\cal M} = \{O_k\}_k$. 
If Alice can perform all possible projective measurements, we denote 
her measurement strategy by $\mathcal{ M}^{A}$. 
In this situation,  EPR-steering can be interpreted as an inclusion problem in convex geometry~\cite{Ngu19}.
This inclusion problem is converted to an optimization problem to compute the critical radius, which serves as a criterion to differentiate between steerable states and non-steerable states.
It states that \emph{a two-qubit state can be used for EPR-steering, if and only if the critical radius $R(\rho_\mathrm{AB})<1$}~\cite{Ngu19}. 
However, deriving analytical results for the critical radius can be challenging. To address this,  Nguyen and co-workers developed a linear program to compute an upper bound $R^U(\rho_\mathrm{AB})$ and a lower bound $R^L(\rho_\mathrm{AB})$ on the critical radius for a given two-qubit state~\cite{Ngu19}. 
The parameter $N_{\mathrm{vert}}$ in the numerical method represents the number of vertices of inner (outer) polytopes used to approximate the Bloch sphere. For states with axial symmetry, the largest parameter used is $N_{\mathrm{vert}}^{\mathrm{max}}=1514$~\cite{Ngu19}. Since the three quantum states under consideration possess axial symmetry, we choose to use the maximum value of $N_{\mathrm{vert}}$ in our calculations to ensure the highest achievable level of accuracy.

In realistic experiments, Alice can only do some finite number of measurements. Here are some cases we consider in this paper. The first one, which is not finite, was already introduced above. 
\begin{enumerate}
\item All projective measurements $\mathcal{M}^{A}$,
\item Two measurements $\mathcal{M}_2=\{\sigma_x, \sigma_z\}$,
\item Three measurements $\mathcal{M}_3=\{\sigma_x, \sigma_y, \sigma_z\}$,
\item $m$ measurements 
$\mathcal{M}^D_m=\{\sigma_z, \sigma_\theta\}_{\theta=l\pi/m}$, \end{enumerate}
where $\sigma_\theta=\cos{(\theta)}\sigma_x+\sin{(\theta)}\sigma_y$  and $l=0,1,2,\cdots,2m-1$. 
This last was a measurement strategy introduced in Ref.~\cite{Jones11}, involving several equatorial observables and one nonequatorial observable, $\sigma_z$. 
Note that this includes the other finite strategies as special cases: $\mathcal{M}^D_1 = \mathcal{M}_2$, and $\mathcal{M}^D_2 = \mathcal{M}_3$. Also, 
$\underset{m\to \infty }{\lim}\mathcal{M}^D_m = \mathcal{M}^E$, where 
\begin{align} 
\mathcal{M}^{E}=\{\sigma_z, \sigma_\omega\}_{\omega \in[0,2\pi)},
\end{align}
where $\sigma_\omega=\cos{(\omega)}\sigma_x+\sin{(\omega)}\sigma_y$. This strategy was introduced in Ref.~\cite{Jon07}.

\section{The steerability of X-states}
\label{sec:The steerability of X-states}

Any two-qubit state $\rho_\mathrm{AB}$ can be expressed in terms of the Pauli matrices as
\begin{equation}\label{atq}
\rho_\mathrm{AB}
=\frac{1}{4}\bigg(\openone \otimes \openone+{\boldsymbol{a}} \cdot \boldsymbol{\sigma} \otimes \openone+\openone\otimes {\boldsymbol{b}} \cdot \boldsymbol{\sigma}
+\!\sum_{i,j=x,y,z}\! T_{i j}\sigma_{i} \otimes \sigma_{j}\bigg),
\end{equation}
where $\boldsymbol{\sigma}=(\sigma_x,\sigma_y,\sigma_z)$ denotes the Pauli matrices,  ${\boldsymbol{a}}=({a}_x,{a}_y,{a}_z)$ and ${\boldsymbol{b}}=({b}_x,{b}_y,{b}_z)$ represents the Bloch vectors for Alice and Bob’s qubits, respectively, and ${T}$ is the spin correlation matrix. In terms of components, 
${a}_{i}=\operatorname{Tr}\left[\rho_\mathrm{AB}(\sigma_{i} \otimes \openone)\right]$, 
${b}_{j}=\operatorname{Tr}\left[\rho_\mathrm{AB}(\openone \otimes \sigma_{j})\right]$, and ${T}_{ij}=\operatorname{Tr}\left[\rho_\mathrm{AB}(\sigma_{i} \otimes \sigma_{j})\right] \ (i, j=x,y,z).$ 

In this paper, we consider three different scenarios in which the associated quantum entangled states are X-states~\cite{Yu07}. In the computational basis $\{|00\rangle,|01\rangle,|10\rangle,|11\rangle\}$, the density matrix of a two-qubit X-states can be expressed in the following form
\begin{align}\label{dm-x-states}
\mathcal{X}=\begin{pmatrix}
\mathcal{X}_{11} & 0 & 0 & \mathcal{X}_{14} \\
0 & \mathcal{X}_{22} & \mathcal{X}_{23} & 0 \\
0 & \mathcal{X}_{23}^\ast & \mathcal{X}_{33} & 0 \\
\mathcal{X}_{14}^\ast & 0 & 0 & \mathcal{X}_{44}
\end{pmatrix},
\end{align}
which has seven independent real parameters.  
The elements of two-qubit X-states can be transformed into real numbers via local unitary transformations that preserve steerability~\cite{Chen11}. Therefore, we only need to consider the following density matrix, which is characterized by five real parameters
\begin{equation}\label{x-states}
\rho_\text{X}=\frac{1}{4}\bigg(\openone \otimes \openone+a \sigma_{z} \otimes \openone+\openone \otimes b \sigma_{z}+\sum_{i=x,y,z} t_{i} \sigma_{i} \otimes \sigma_{i}\bigg),
\end{equation}
where $\boldsymbol{a}=(0,0,a)$, $\boldsymbol{b}=(0,0,b)$ and ${T}=\operatorname{diag}\left(t_{x}, t_{y}, t_{z}\right)$.
To understand X-states better in the context of quantum information, one can refer to Ref.~\cite{Rau09}, which provides an algebraic characterization. For X-states, the concurrence is given by~\cite{Wang06}
\begin{align}\label{concurrence}
C(\mathcal{X})=2\max[0, |\mathcal{X}_{32}|-\sqrt{\mathcal{X}_{44}\mathcal{X}_{11}}, |\mathcal{X}_{41}|-\sqrt{\mathcal{X}_{33}\mathcal{X}_{22}}].
\end{align}
The concurrence quantifies entanglement and is zero if and only if the state is separable~\cite{Wootters98}.

In Ref.~\cite{Jon07}, the authors considered the measurement strategy $\mathcal{M}^E$, defined in Sec.~\ref{sec:EPR-steering}, to find a sufficient condition for the steerability of a class of states called inept states. These are a sub-class of $X$-states. Here we apply the same method to establish a sufficient condition for the steerability of X-states with restriction $|t_x|=|t_y|$,  as per the following Theorem. 
\begin{theorem}\label{Theorem}
\!. For two-qubit X-states with $|t_x|=|t_y|=: t_\perp$ shared by Alice and Bob, Alice can demonstrate steering using the measurement strategy $\mathcal{M}^D_m$ if and only if
\begin{align}\label{xsteeringineqnnn}
&2 m \sin \left(\frac{\pi }{2 m}\right)>\frac{F}{t_\perp},
\end{align}
where
\begin{equation}
F=\sqrt{(1+a)^2-\left(b+t_z\right)^2}+\sqrt{(1-a)^2-\left(b-t_z\right)^2}.
\end{equation}
\end{theorem}

\textit{Proof}. See Appendix~\ref{sec:appendixa} for the proof of this Theorem.\\

It is easy to calculate that $\underset{m\to \infty }{\lim} m\sin \left(\frac{\pi }{2 m}\right)=\frac{\pi }{2}$. 
Hence we have the following  corollary. 
\begin{corollary}\label{Corollary}
\!. For two-qubit X-states with $|t_x|=|t_y| =: t_\perp$ shared by Alice and Bob, if Alice uses the measurement strategies $\mathcal{M}_2$, $\mathcal{M}_3$, and $\mathcal{M}^{E}$,
then Alice can demonstrate steering if and only if
\begin{align}
2>&\frac{F}{t_\perp},\label{xsteeringc2}\\
2\sqrt{2}>&\frac{F}{t_\perp},\label{xsteeringc3}\\
\pi>&\frac{F}{t_\perp},\label{xsteeringce}
\end{align}
respectively.
\end{corollary}

\section{Steering in multi-party two-producible states}
\label{sec:Multiparty Steering}

A pure state $|\varphi\rangle$ of a quantum system of $n$ parties is termed $k$-party entangled (or $k$-producible, for short) if it can be written as $|\varphi\rangle=|\phi_1\rangle\otimes|\phi_2\rangle\otimes\cdots\otimes|\phi_f\rangle$, where the states $|\phi_i\rangle$ for all $i=1,2,\cdots,f$ are states of maximally $k$ parties~\cite{guhne05}. A mixed state is $k$-producible if it can be written as a mixture of $k$-producible pure states. The correlation of a $k$-producible pure (mixed) state cannot be produced by $(k-1)$-producible entanglement~\cite{guhne05}. The notion of producibility is also termed the depth of entanglement~\cite{Sorensen01}. In this paper, we consider 2-producible $n$-partite states.

Consider a multipartite ($n>2$) qubit state in which the entanglement is produced by mixing the following resources: first, two-qubit non-maximally entangled pure states 
\begin{align}\label{entstate}
\rho_\alpha=|\Psi_\alpha\rangle\langle\Psi_{\alpha}|,
\end{align}
with (for $0 < \alpha \leq 1/2$) 
\begin{align}
|\Psi_{\alpha}\rangle=
\sqrt{1-\alpha}|0\rangle |0\rangle+\sqrt{\alpha}|1\rangle |1\rangle;
\end{align}
and second, single-qubit pure states 
\begin{equation}
    \rho_{0}=\ket{0}\bra{0}. \label{phi}
\end{equation}
We use these according to three distinct production scenarios:  

\emph{Scenario 1 (Random $\frac{n}{2}$-Pair Entanglement (R$\frac{n}{2}$PE))}. Company $C_1$ manufactures $\frac{n}{2}$ identical entangled qubit pairs as in Eq.~\eqref{entstate}, where $n$ is an even number greater than 2. It randomly delivers the $n$ qubits to $n$ parties, $P_i$  $(i=1,2,\cdots,n)$, with each party receiving exactly one qubit. 

\emph{Scenario 2 (Random Pair Entanglement (RPE))}.  Company $C_2$ manufactures only \emph{one} entangled pair of qubits as in Eq.~\eqref{entstate}, and $n-2$ independent qubits as in Eq.~\eqref{phi}. It randomly delivers  
the $n$ qubits to the $n$ parties, as per company $C_1$.  
That is, two random parties share the entangled state \eqref{entstate}, while each of other parties has exactly one pure qubit~\eqref{phi}. 

\emph{Scenario 3 (Semi-Random Pair Entanglement (SRPE))}  Company $C_3$, as per company $C_2$, manufactures only one entangled pair of qubits as in Eq.~\eqref{entstate} and $n-2$ independent qubits as in Eq.~\eqref{phi}. It delivers one qubit from the entangled pair to a special party $A$ (Alice), and it randomly delivers the remaining $n-1$ qubits  to the remaining $n-1$ parties  $B_j$ $(j=1,2,\cdots,n-1)$ (Bobs), such that each receives exactly one qubit. each. That is, only Alice and one random Bob share the entangled state~\eqref{entstate}, while the other $n-2$ Bobs each have exactly one pure qubit~\eqref{phi}.   

Later, we will also consider the case where the entangled quantum state $\rho_{\alpha}$ is contaminated with noise of magnitude $\mu\in[0,1]$, as here: 
\begin{align}\label{entstatenoise}
\rho^{\mu}_{\alpha}=(1-\mu)\rho_{\alpha}
+\mu \frac{\openone_{2\times2}\otimes \openone_{2\times2}}{4}.
\end{align}

\subsection{Random $\frac{n}{2}$-Pair Entanglement (R$\frac{n}{2}$PE)}
\label{sec:Rn2PE}

\begin{figure}[h]\centering
\includegraphics[angle=0,width=0.16\linewidth]{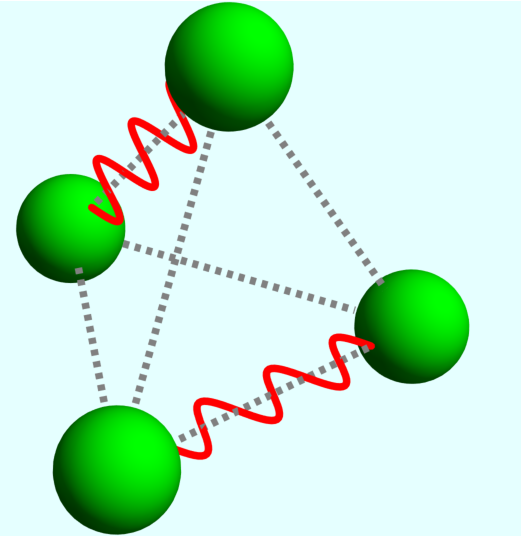}
\hspace{0.1cm}
\includegraphics[angle=0,width=0.16\linewidth]{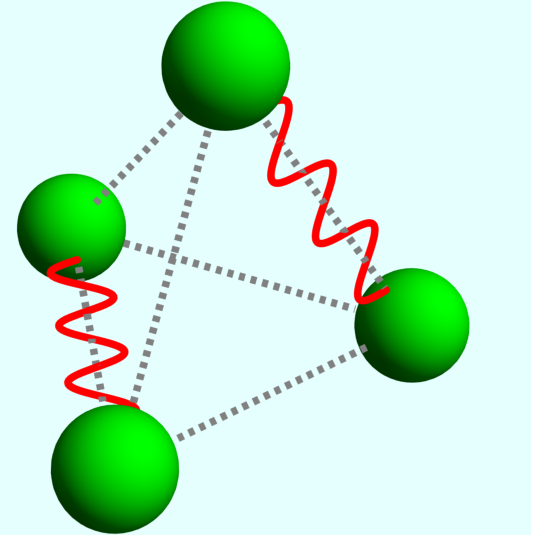}
\hspace{0.1cm}
\includegraphics[angle=0,width=0.16\linewidth]{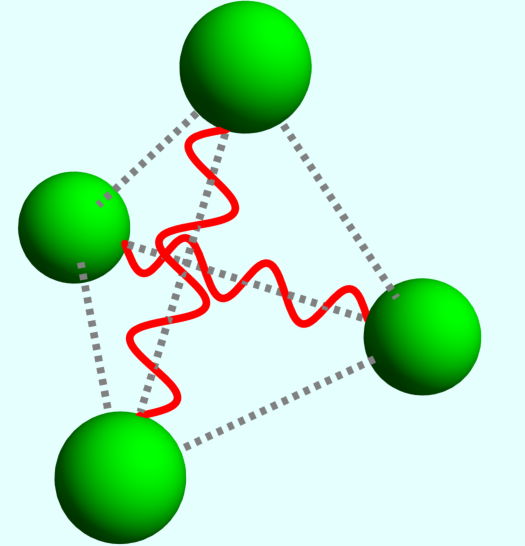}
\hspace{0.2cm}
\includegraphics[angle=0,width=0.15\linewidth]{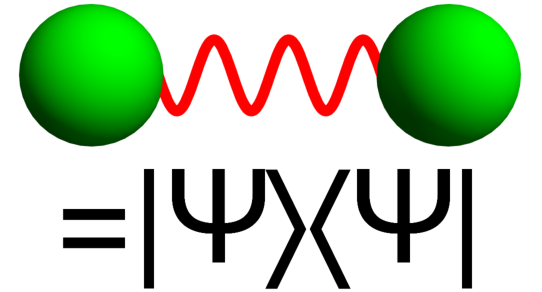} 
\caption{R$\frac{n}{2}$PE. e.g. $n=4$. Each green ball represents a party, while the wavy line denotes the entangled state $|\Psi_{\alpha}\rangle$ shared by two parties. The overall state is an equal mixture of the three cases shown.}\label{fig-half}
\end{figure}

R$\frac{n}{2}$PE is the scenario that the ensemble is composed of $\frac{n}{2}$ pairs of parties sharing the entangled state $|\Psi_{\alpha}\rangle$ in Eq.~\eqref{entstate}. Here, $n$ is an even number greater than 2. In Fig.~\ref{fig-half}, we present the case when $n=4$. 

A permutation of $\{1,2,\cdots,n\}$ is a one-to-one function $\tau:\{1,2,\cdots,n\}\to\{1,2,\cdots,n\}$. There are $n!$ different permutations, and the set of all such permutations forms a group with respect to the composition of functions~\cite{horn2012matrix}. Using this representation, $n$-partite random $\frac{n}{2}$-pair entangled (R$\frac{n}{2}$PE) state can be written as
\begin{align}
\rho_{\mathrm{R}\frac{n}{2}\mathrm{PE}}^n=\frac{1}{n!}\sum_{\tau}\overbrace{\rho_{\tau(1)\tau(2)}\otimes \cdots \otimes\rho_{\tau(n-1)\tau(n)}}^{n/2\ \mathrm{times}}
\label{eq:rn2pe_def}
\end{align}
in which the sum is over all permutations and for each pair shown $\rho_{\tau(i)\tau(j)}=\rho_\alpha$ which is defined in Eq.~\eqref{entstate}.
Based on this, the bipartite reduced R$\frac{n}{2}$PE state shared by any two parties (Alice and Bob) can be written in the following form
\begin{align} \label{bip1}
\varrho_{\mathrm{R}\frac{n}{2}\mathrm{PE}}^{2\mathrm{r}}=S\rho_{\alpha}+(1-S)\varrho_1\otimes\varrho_1,
\end{align}
where  
$\varrho_1= {\rm Tr}_\mathrm{A}[\rho_{\alpha}]= {\rm Tr}_\mathrm{B}[\rho_{\alpha}]$ and $S={1}/({n-1})$.  
The above state is a special case of the family of so-called inept states~\cite{Jones05,Jon07}, which can have any value for $S$.
Interestingly, the R$\frac{n}{2}$PE scenario can also be interpreted in the context of a perfect matching problem in graph theory \cite{bondy1976graph}.
That is, if each of the $n$ parties are treated as vertices in a graph, and each entangled state corresponds to an edge, all perfect matchings in such define the terms inside the sum in Eq.~\eqref{eq:rn2pe_def}.
An equal mixture of all such perfect matchings gives the global R$\frac{n}{2}$PE quantum state.
When considering the case with noise $\mu$, the bipartite reduced R$\frac{n}{2}$PE state becomes
\begin{align}
\varrho_{\mathrm{R}\frac{n}{2}\mathrm{PE}}^{2\mathrm{r}\mu}=\frac{1}{n-1}\rho^{\mu}_\alpha
+\frac{n-2}{n-1}\varrho_1^{\mu}\otimes\varrho_1^{\mu},
\end{align}
where $\varrho_1^{\mu}={\rm Tr}_\mathrm{A}[\rho^\mu_\alpha]={\rm Tr}_\mathrm{B}[\rho^\mu_\alpha]$ and $\rho^\mu_\alpha$ is defined in Eq.~\eqref{entstatenoise}.  For the other two scenarios, there are similar expressions with noise $\mu$ that replace $\rho_\alpha$, $\varrho_1$ with $\rho_\alpha^\mu$, $\varrho_1^\mu$, respectively.

\begin{figure}[h]\centering
\includegraphics[angle=0,width=0.87\linewidth]{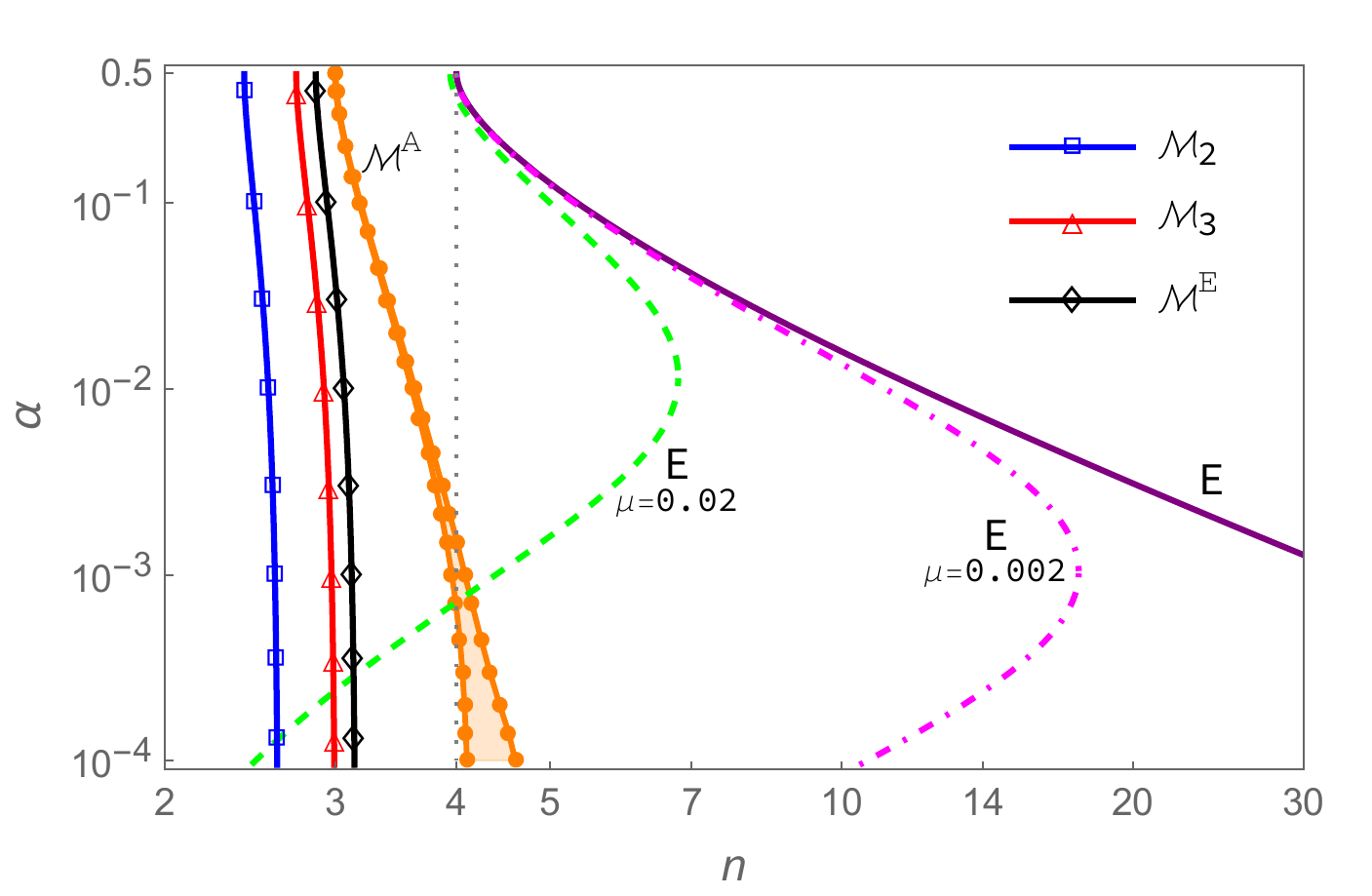}
\caption{The steerability of the bipartite reduced R$\frac{n}{2}$PE state. 
The left-hand side region of each curve represents the associated entangled or steerable region. 
The dashed green, dot-dashed magenta and purple curves denote the entanglement ($\mathsf{E}$) bounds with noise $\mu = 0.02, 0.002, 0$, respectively. The blue (square markers), red (triangle markers), and black (diamond markers) lines represent the steering bounds for measurement schemes $\mathcal{M}_2$, $\mathcal{M}_3$, and $\mathcal{M}^{E}$ respectively.  The orange joined dots denote the steering (lower and upper) bounds without noise for all projective measurements $\mathcal{M}^{A}$. The shaded region between upper and lower bound is the states for which the numerical imprecision prevented a classification as steerable or unsteerable. In the plot we treat $n$ as a real parameter for simplicity, although physically it must be an even integer.} \label{steering-half} 
\end{figure}

Using Eq.~\eqref{concurrence}, we find the entanglement condition for  Eq.~\eqref{bip1}, the bipartite reduced R$\frac{n}{2}$PE state (without noise), to be 
\begin{align}
1-2\alpha>\frac{\sqrt{n(n-4)}}{n-2}.
\end{align}
This is plotted in Fig.~\ref{steering-half} along with the entanglement bounds with noise $\mu=0.02,0.002$. 

For the steerability of bipartite reduced R$\frac{n}{2}$PE state. 
According to Theorem~\ref{Theorem}, if Alice makes two, three and equatorial measurements $\mathcal{M}_2$, $\mathcal{M}_3$ and $\mathcal{M}^{E}$, the left-hand sides of Eqs.~(\ref{xsteeringc2}-\ref{xsteeringce}) remain unchanged, and their right-hand sides become
\begin{equation}
\frac{F}{ t_\perp}=
2\sqrt{n-2}\left(\sqrt{\alpha\eta}+\sqrt{(1-\alpha)(n-\eta)}\right).
\end{equation}
where $\eta:=1+\alpha(n-2)$.
If Alice makes all projective measurements $\mathcal{M}^{A}$, 
we can determine the steering bound of bipartite reduced R$\frac{n}{2}$PE state by calculating its critical radius with a numerical algorithm~\cite{Ngu19}, as shown in Fig.~\ref{steering-half}. Note the large region of parameter space where the state is entangled but steering is not possible. In particular, entanglement can exist for arbitrarily large $n$ by making $\alpha$ small, while 
steering is only shown to be possible up to $n=4$. It seems unlikely that it would be possible for $n=6$, no matter how small $\alpha$ is. (Recall that $n$ must be even for this scenario.)

\subsection{Random Pair Entanglement (RPE)}

\begin{figure}[h]\centering
\includegraphics[angle=0,width=0.15\linewidth]{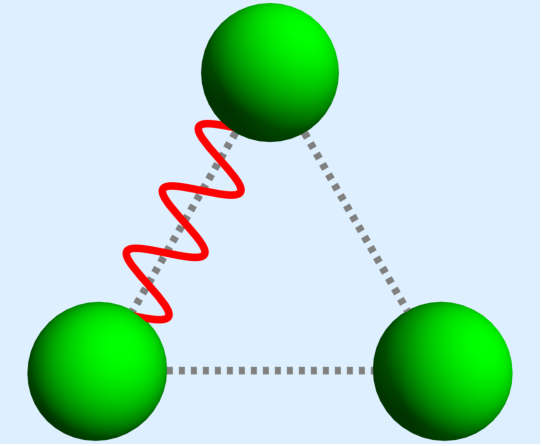}
\hspace{0.05cm}
\includegraphics[angle=0,width=0.15\linewidth]{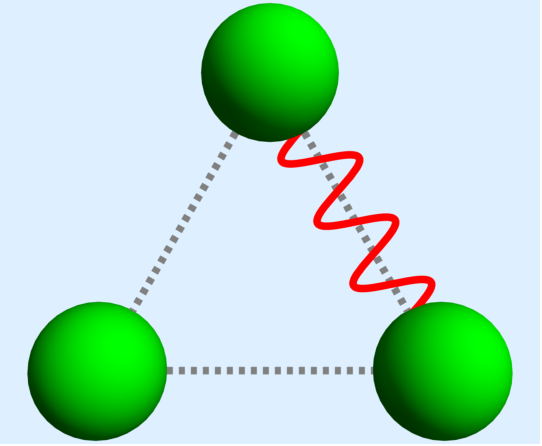}
\hspace{0.05cm}
\includegraphics[angle=0,width=0.15\linewidth]{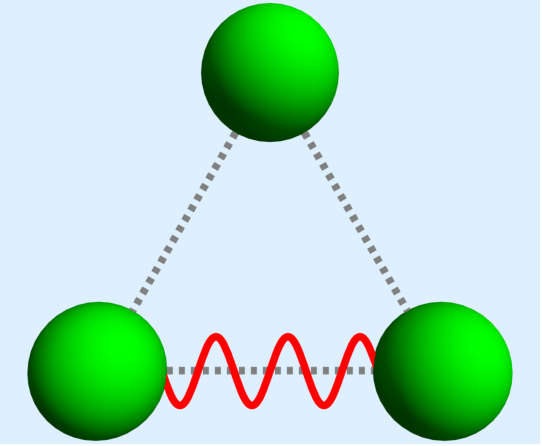}\\
\vspace{0.3cm}
\includegraphics[angle=0,width=0.15\linewidth]{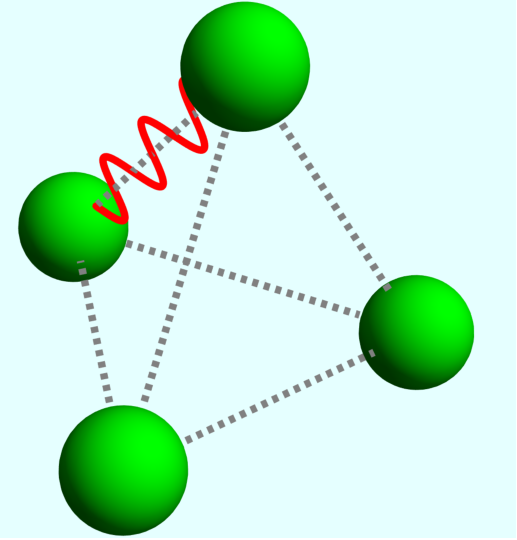}
\hspace{0.0cm}
\includegraphics[angle=0,width=0.15\linewidth]{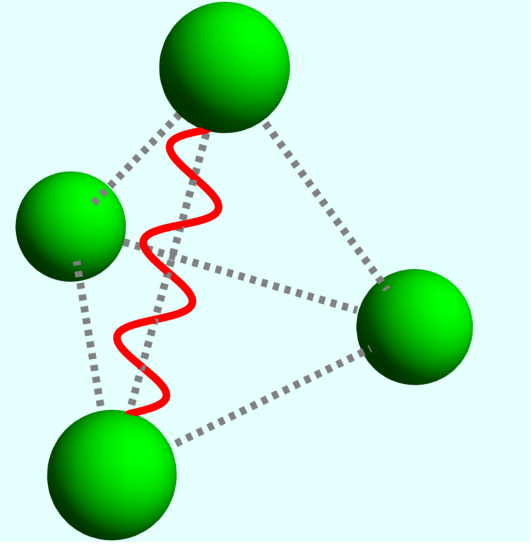}
\hspace{0.0cm}
\includegraphics[angle=0,width=0.15\linewidth]{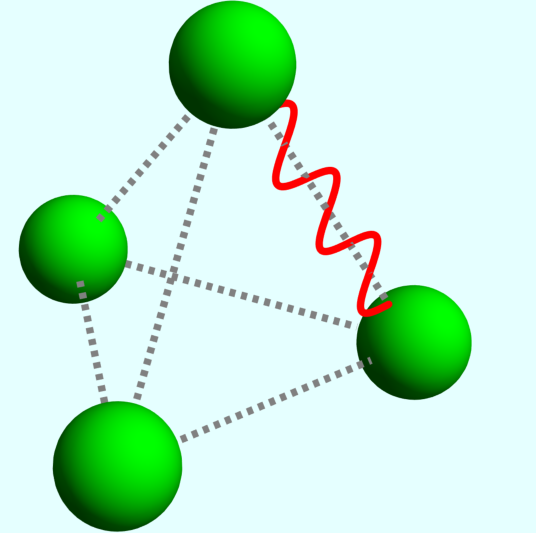}
\hspace{0.0cm}
\includegraphics[angle=0,width=0.15\linewidth]{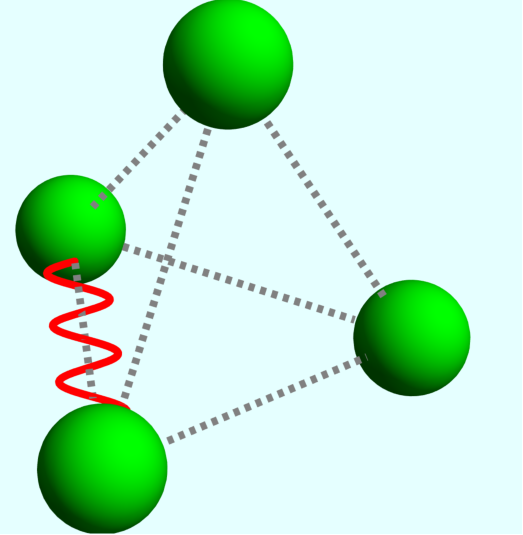}
\hspace{0.0cm}
\includegraphics[angle=0,width=0.15\linewidth]{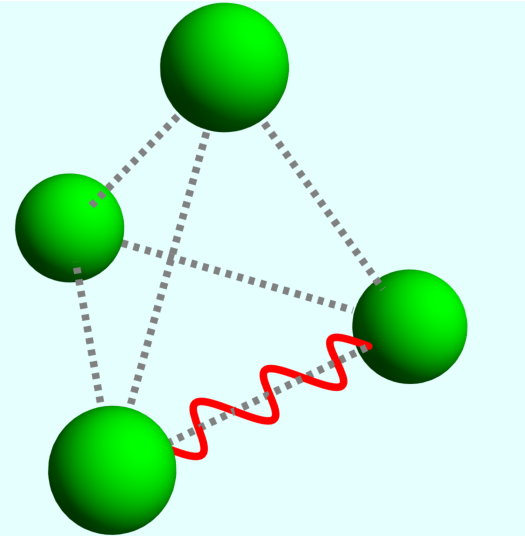}
\hspace{0.0cm}
\includegraphics[angle=0,width=0.15\linewidth]{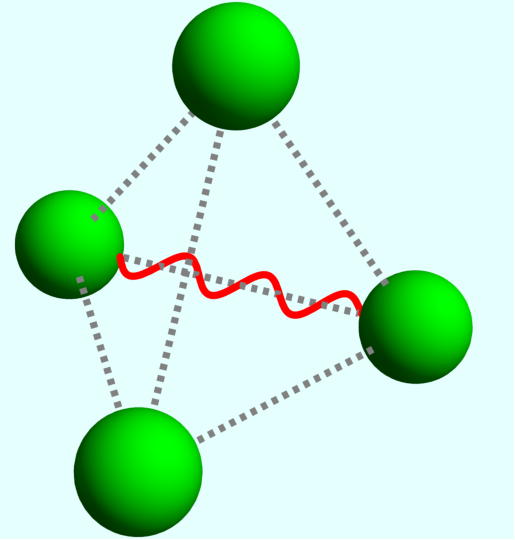}
\caption{RPE state for $n=3$ and $4$. Only a random pair of parties among the $n$ parties shares an entangled state.   
The overall state is an equal mixture of the three and six cases shown respectively.
}\label{fig-rpe}
\end{figure}
RPE is the scenario that only a random pair of parties among $n$ parties shares the entangled state $|\Psi_{\alpha}\rangle$ and other parties have the single-qubit pure state  $|0\rangle$. 
Fig.~\ref{fig-rpe} represents the examples of $n=3,4$. The $n$-partite RPE state is of the form
\begin{align}
\rho_\mathrm{RPE}^n=\frac{1}{n(n-1)}\sum_{1\leq i<j\leq n}(\rho_{ij}+\rho_{ji})\otimes\rho_{0}^{\otimes(n-2)},
\end{align}
where $\rho_{ij}=\rho_\alpha$ defined in Eq.~\eqref{entstate}.
The bipartite reduced RPE state can be written as
\begin{align}
\varrho_{\rm RPE}^{2\mathrm{r}}=
&\frac{2}{n(n-1)}\rho_{\alpha}
+\frac{2(n-2)}{n(n-1)}\varrho_1\otimes\rho_{0}\\
&+\frac{2(n-2)}{n(n-1)}\rho_{0}\otimes\varrho_1
+\frac{n^{2}-5n+6}{n(n-1)}\rho_{0}\otimes\rho_{0}.\notag
\end{align}

\begin{figure}[h]\centering
\includegraphics[angle=0,width=0.87\linewidth]{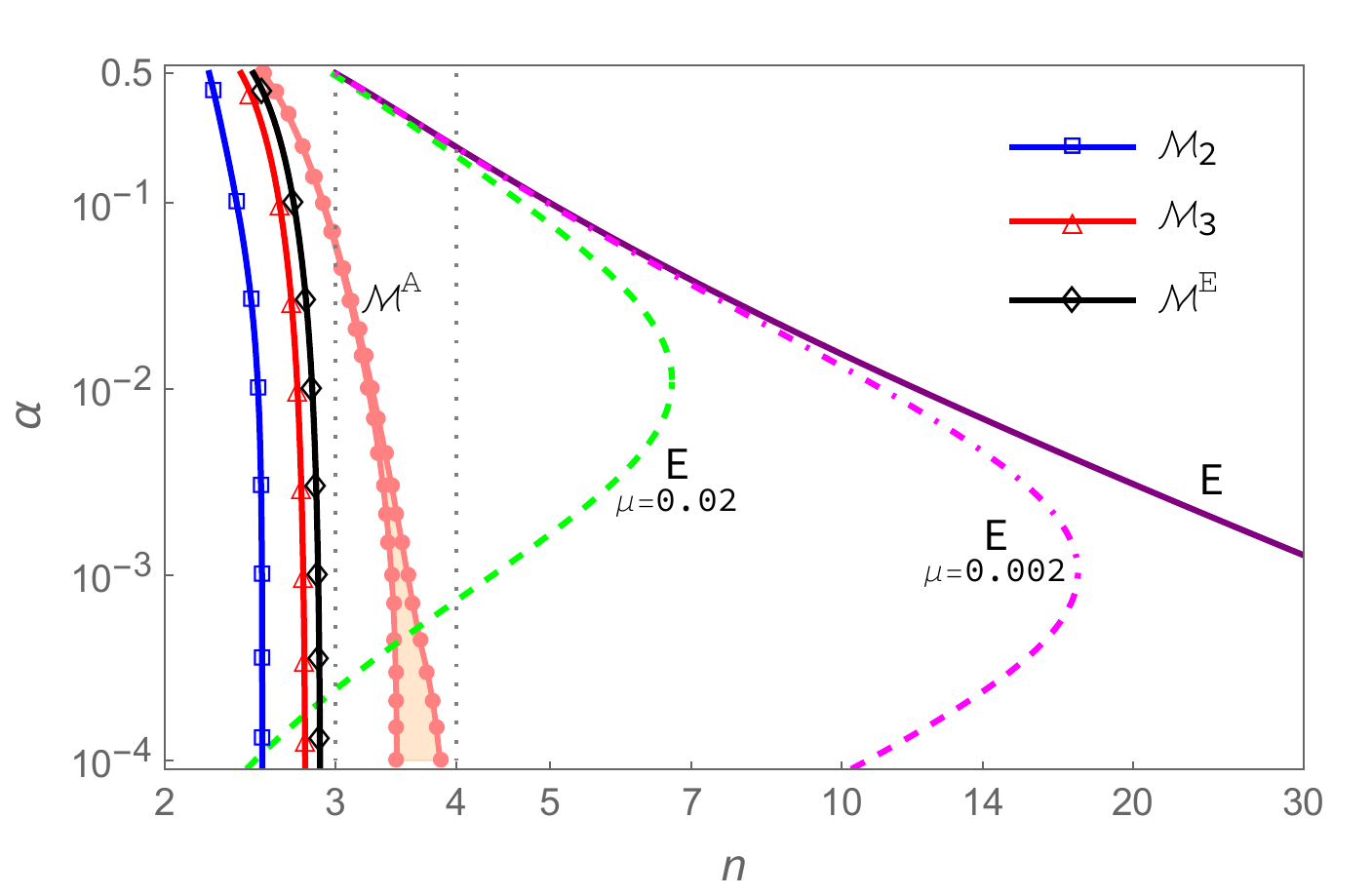}
\caption{The steerability of the bipartite reduced RPE state. 
The region to the left of every curve illustrates the associated entangled or steerable region. 
The dashed green, dot-dashed magenta and purple curves denote the entanglement ($\mathsf{E}$) bounds with noise $\mu = 0.02, 0.002, 0$, respectively. The blue (square markers), red (triangle markers), and black (diamond markers) lines represent the steering bounds for measurement schemes $\mathcal{M}_2$, $\mathcal{M}_3$, $\mathcal{M}^{E}$, respectively. The orange joined dots denote the steering (lower and upper) bounds without noise for all projective measurements $\mathcal{M}^{A}$. The shaded area contains states not classified as either steerable or non-steerable by the program. For ease of understanding in the plot, we consider the variable $n$ as a real parameter, although it must be an integer in physical terms.} \label{steering-rpe} 
\end{figure}
The entanglement criterion of bipartite reduced RPE state without noise reads
\begin{align}
\alpha<\frac{1}{n^2-4 n+5}.
\end{align}
The entanglement bounds with noise $\mu=0.02,0.002,0$ are shown in Fig.~\ref{steering-rpe}. 
For the steerability of bipartite reduced RPE state, we again apply Theorem~\ref{Theorem}. if Alice makes two, three and equatorial measurements $\mathcal{M}_2$, $\mathcal{M}_3$ and $\mathcal{M}^{E}$, the right hand sides of Eqs.~(\ref{xsteeringc2}-\ref{xsteeringce}) become 
\begin{equation}
\frac{F}{t_\perp}=\frac{\sqrt{2 (n-2)} \left(\sqrt{2 \alpha}+\sqrt{6 \alpha+n (n-1-4 \alpha) }\right)}{\sqrt{1-\alpha }}.
\end{equation}
If Alice performs all projective measurements $\mathcal{M}^{A}$, the upper and lower steering bounds can be computed by the numerical algorithm~\cite{Ngu19}. These bounds are all shown in Fig.~\ref{steering-rpe}. The behaviour is very similar to that of the R$\frac{n}{2}$PE states in Fig.~\ref{steering-half}, but here $n=3$ is the largest certified steerable state for any $\alpha$, and it seems unlikely that $n=4$ would allow steering.

\subsection{Semi-Random Pair Entanglement (SRPE)}
\begin{figure}[h]\centering
\includegraphics[angle=0,width=0.16\linewidth]{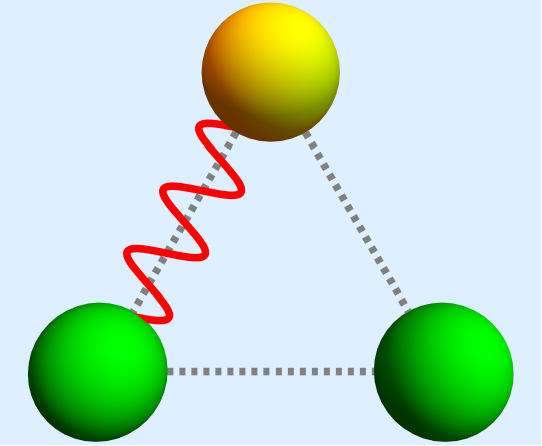}
\hspace{0.05cm}
\includegraphics[angle=0,width=0.16\linewidth]{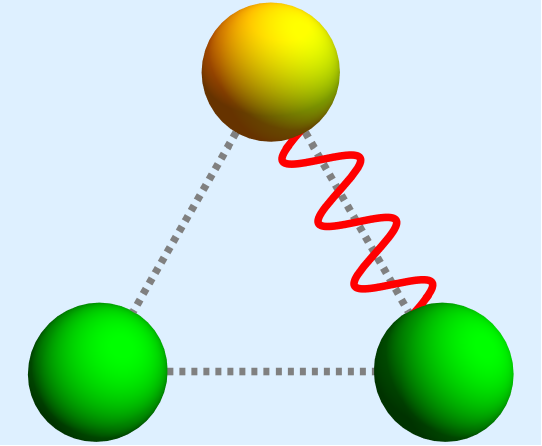}\\
\vspace{0.3cm}
\includegraphics[angle=0,width=0.16\linewidth]{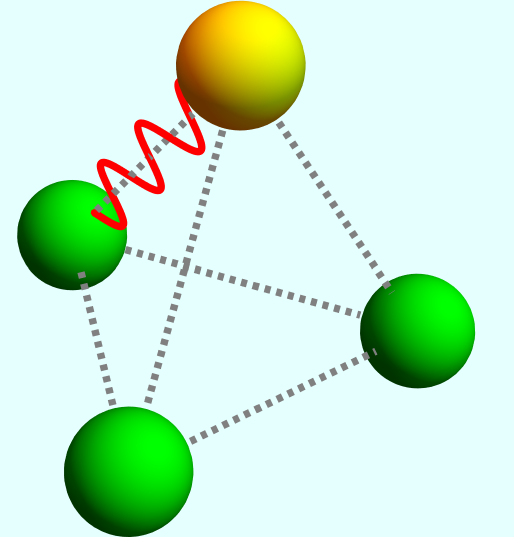}
\hspace{0.05cm}
\includegraphics[angle=0,width=0.16\linewidth]{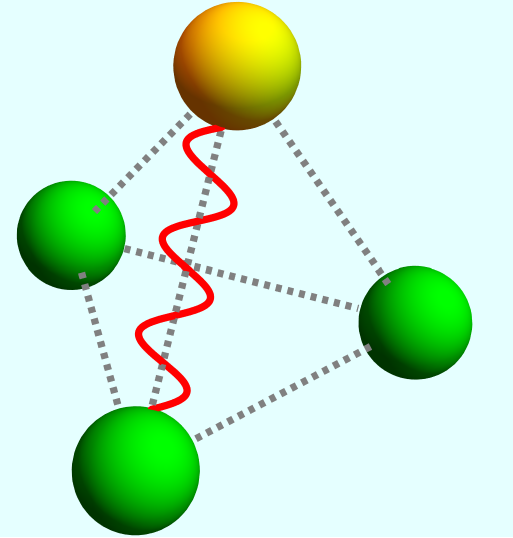}
\hspace{0.05cm}
\includegraphics[angle=0,width=0.16\linewidth]{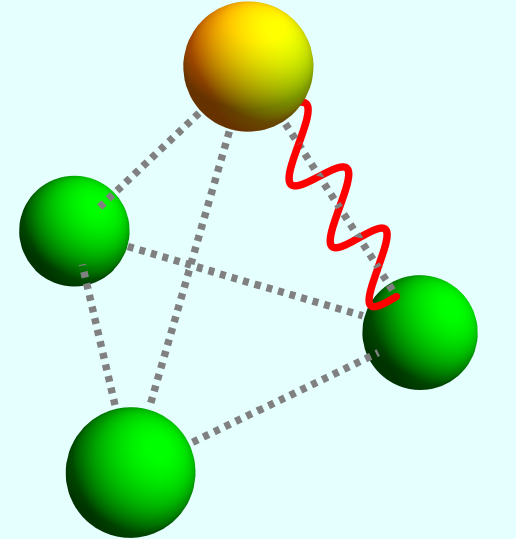}
\caption{SRPE. e.g. $n=3,4$. The green ball represents an arbitrary party. The yellow denotes represents the fixed party, which randomly shares the entangled state $|\Psi_{\alpha}\rangle$ with an arbitrary party.}\label{fig-srpe}
\end{figure}

\begin{figure}[h]\centering
\includegraphics[angle=0,width=0.87\linewidth]{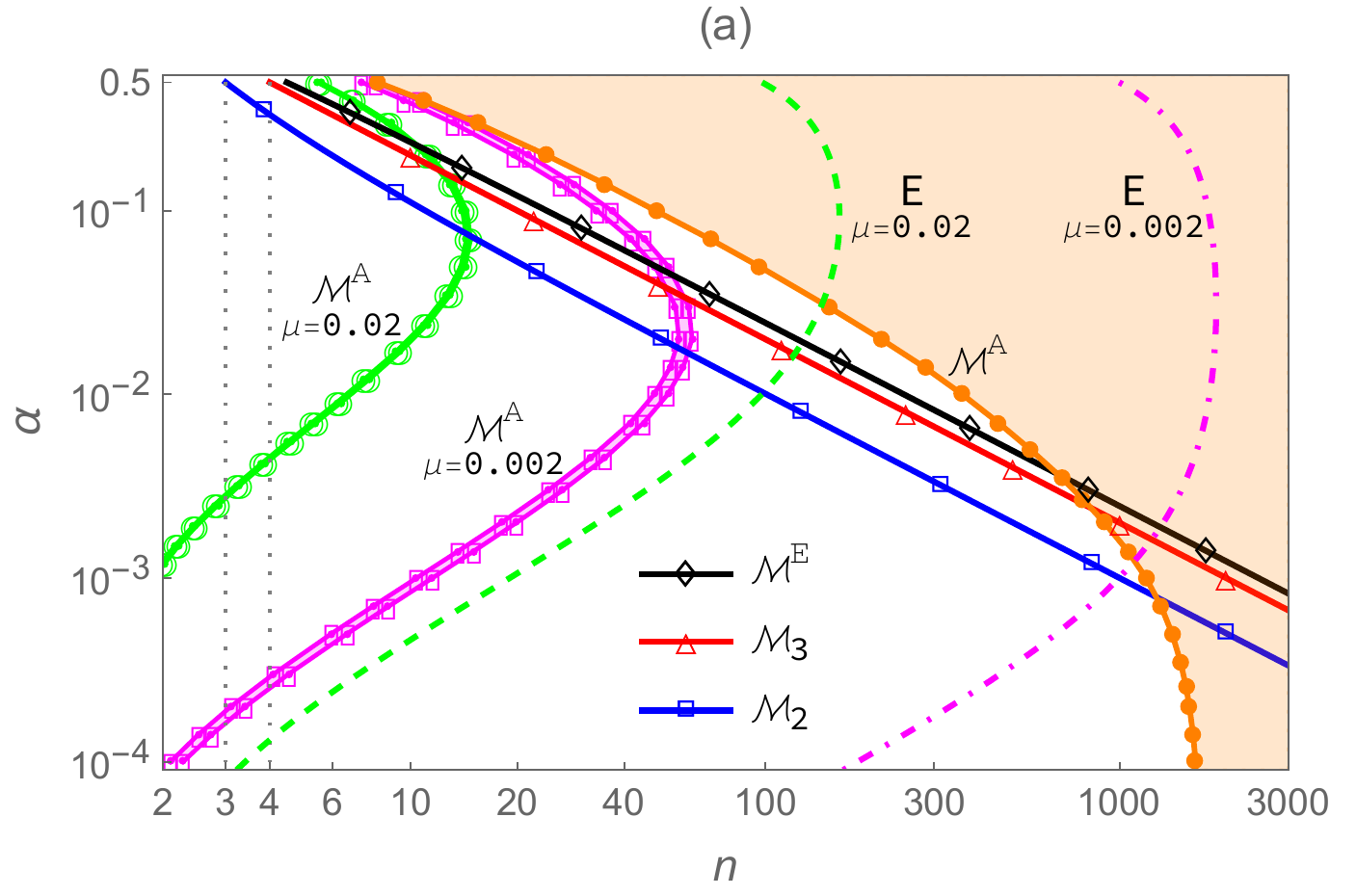}\\
\vspace{0.1cm}
\includegraphics[angle=0,width=0.87\linewidth]{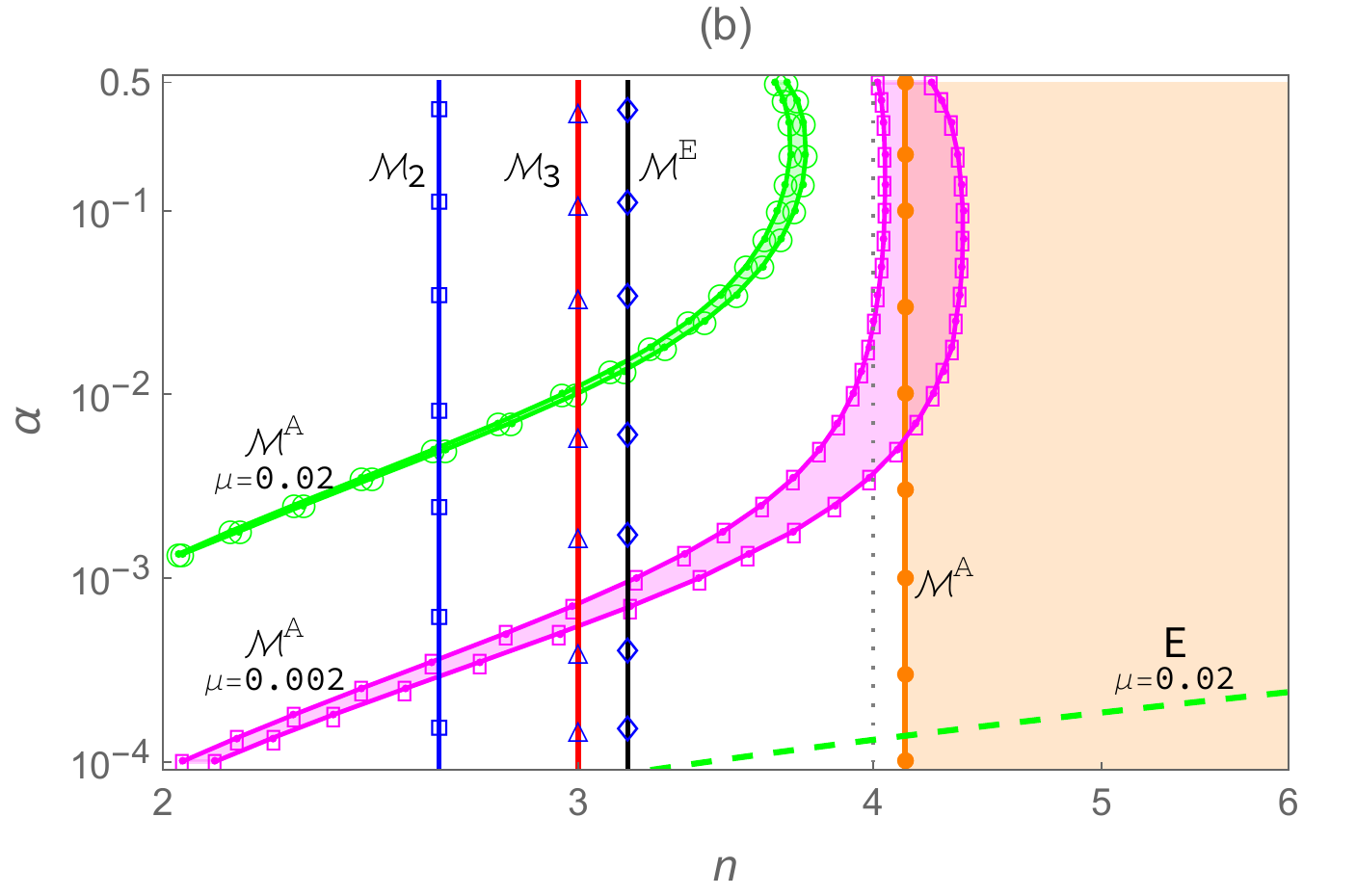}
\caption{The steerability of the bipartite reduced SRPE state. (a) Steering from Alice to Bob. (b) Steering from Bob to Alice. 
The region to the left of every curve signifies its respective entangled or steerable region. 
The dashed green and dot-dashed magenta curves denote the entanglement ($\mathsf{E}$) bounds with noise $\mu = 0.02, 0.002$, respectively. The blue (square markers), red (triangle markers), and black (diamond markers) lines represent the steering bounds for measurement schemes $\mathcal{M}_2$, $\mathcal{M}_3$, $\mathcal{M}^{E}$, respectively. The green (circle markers) and magenta (rectangle markers) joined dots denote the steering bounds (including lower bound and upper bound) with noise $\mu = 0.02, 0.002$ for all projective measurements $\mathcal{M}^{A}$, respectively. The orange joined dots denote the steering lower bound without noise for all projective measurements $\mathcal{M}^{A}$, but the upper bound cannot be found by the numerical method. The shaded region corresponds to the states where numerical imprecision prevents distinguishing whether the state is steerable or non-steerable.} \label{steering-srpe} 
\end{figure}

SRPE is the scenario that among $n$ parties a fixed party shares the entangled state $|\Psi_{\alpha}\rangle$ with a random party and other parties are prepared in single-qubit pure state $|0\rangle$ each. Fig.~\ref{fig-srpe} illustrates the examples $n=3,4$. The $n$-partite semi-random pair entangled (SRPE) state can be written as
\begin{align}
\rho_\mathrm{SRPE}^n=\frac{1}{n-1}\sum_{j=1}^{n-1}\rho_{AB_{j}}\otimes\rho_{0}^{\otimes(n-2)},
\end{align}
where $\rho_{AB_{j}}=\rho_\alpha$.
This gives the bipartite reduced SRPE state
\begin{align}
\varrho_{\rm SRPE}^{2\mathrm{r}}=\frac{1}{n-1}\rho_{\alpha}
+\frac{n-2}{n-1}\varrho_1\otimes\rho_{0}.
\end{align}
From Eq.~\eqref{concurrence}, the above state is entangled for all $\alpha \in (0,1/2]$. The  calculated entanglement bounds with noise $\mu=0.02,0.002$ are shown in Fig.~\ref{steering-srpe}(a). These bounds show a maximum range of $n$ for which entanglement exists, which is reached for some optimal value of $\alpha$. This is similar to the preceding scenarios, but here the optimal value of $\alpha$ is considerably larger. 

Unlike the bipartite reduced R$\frac{n}{2}$PE and RPE states, the bipartite reduced SRPE state is asymmetric with respect to Alice and Bob. Therefore, its steerability will also be asymmetric. First, we discuss the steerability of SRPE state from Alice to Bob. 

According to Theorem~\ref{Theorem} and Corollary~\ref{Corollary}, the bipartite reduced SRPE state is steerable, given Alice performs
\begin{enumerate}
\item measurements $\mathcal{M}_2$, if and only
if 
\begin{align}
\alpha<\frac{1}{n-1}.
\end{align}
\item measurements $\mathcal{M}_3$, if and only
if 
\begin{align}
\alpha<\frac{2}{n}.
\end{align}
\item measurements $\mathcal{M}^{E}$, if and only
if 
\begin{align}
\alpha<\frac{\pi ^2}{4(n-2)+\pi ^2}.
\end{align}
\end{enumerate}
Thus, unlike in the preceding scenarios, for the SRPE scenario Alice is able to steer the Bobs for arbitrarily large $n$, by choosing $\alpha$ sufficiently small. That is, steering by Alice of many Bobs is helped by having only a small amount of entanglement in the single entangled pair which is used to produce the multipartite SRPE state.

If Alice makes all projective measurements, scheme $\mathcal{M}^{A}$, we find it is not possible to get the steering upper bound of bipartite reduced SRPE state $\varrho_{\rm SRPE}^{2\mathrm{r}}$ without noise using the numerical algorithm in Ref.~\cite{Ngu19}. Hence we only calculate the steering lower bound. As shown in Fig.~\ref{steering-srpe}, this lower bound intersects the other three bounds for measurement schemes $\mathcal{M}_2$, $\mathcal{M}_3$, $\mathcal{M}^{E}$, for large $n$, which means that the true bound is definitely above the lower bound. We conjecture that the true bound for $\mathcal{M}^{A}$ is asymptotically proportional to $1/n$, like the bounds from the other measurement schemes.  Then we calculate the critical radius $R(\varrho_{\rm SRPE}^{2\mathrm{r}\mu})$ of the state with two different noise $\mu=0.02,0.002$, as shown in Fig.~\ref{steering-srpe} (a). With decreasing noise, the region in which the states can be steerable becomes progressively larger, also resulting in an increase in the size of the uncertainty interval between upper and lower bounds. Similar to the entanglement calculation, there is an optimal $\alpha$ for achieving the greatest range of steering, which decreases with noise.

Now consider steerability of the bipartite reduced SRPE state from Bob to Alice under different measurement strategies. According to Theorem~\ref{Theorem} and Corollary~\ref{Corollary}, the bipartite reduced SRPE state is steerable, given Bob makes
\begin{enumerate}
\item measurements $\mathcal{M}_2$, if and only if
\begin{align}
n<\frac{3+\sqrt{5}}{2}\approx2.61803.
\end{align}
\item measurements $\mathcal{M}_3$, if and only if
\begin{align}
n<3.
\end{align}
\item measurements $\mathcal{M}^{E}$, if and only if
\begin{align}
n<\frac{1}{2}\left(3+\sqrt{1+\pi^2}\right)\approx3.14845.
\end{align}
\end{enumerate}
If Bob makes all projective measurements $\mathcal{M}^{A}$, using the numerical algorithm in Ref.~\cite{Ngu19}, we calculate steering bounds of the bipartite reduced SRPE state with two different noise $\mu=0.02,0.002$ as shown in Fig.~\ref{steering-srpe} (b). For $\mu=0$, we can only obtain the steering lower bound.

It is worth noting that, unlike steering from Alice to Bob, the steering criteria from Bob to Alice for different measurement strategies are independent of $\alpha$. This phenomenon also appears in the two-qutrit partially entangled states~\cite{Qiang22}.  Furthermore, while the state is entangled for all values of $n$, it is only steerable from Bob to Alice when $n$ assumes small values.

For the steerability of bipartite reduced SRPE states with noise $\mu$, we have computed the cases ($\mu$=0.02, 0.002) for all projective measurements, $\mathcal{M}^{A}$, in Fig.~\ref{steering-srpe} numerically.
For the other classes of measurements we consider, we can use Corollary~\ref{Corollary} to give closed-form expressions for the inequalities which can be used to determine steering when these states are noisy.
When Alice performs measurements from $\mathcal{M}_2$, $\mathcal{M}_3$ and $\mathcal{M}^{E}$, the left-hand sides of Eqs.~(\ref{xsteeringc2}-\ref{xsteeringce}) remain the same, while the right-hand sides transform into
\begin{align}
\frac{F}{ t_\perp}
=\frac{\sqrt{AB}+\sqrt{\mu C}}{D},
\end{align}
where
\begin{align}
    A&=\mu+4 \alpha  (1-\mu),\\
    B&=4 \alpha  (1-\mu) (n-2)+\mu  (2 n-3),\\
    C&=4(n-1)(1-\alpha(1-\mu))-\mu(2n-1),\\
    D&=2 (1-\mu) \sqrt{\alpha (1-\alpha)}.
\end{align}
For steering from Bob to Alice, we only need to exchange $a$ and $b$ in Theorem~\ref{Theorem} and Corollary~\ref{Corollary}. 
Using this fact, we can immediately deduce that the right-hand sides of Eqs.~(\ref{xsteeringc2}-\ref{xsteeringce}) become
\begin{align}
\frac{F}{ t_\perp}
=\frac{\sqrt{\mu A}+\sqrt{BC}}{D}.
\end{align}

\section{Steering properties in small networks}
\label{sec:networks}

\begin{figure}[h]
\begin{tabular}{|M|M|M|M|}
\hline
{Scenario} & {Condition} & $n=3$  & $n=4$ \\
\hline
\multirow{8}{*}{\makecell{R$\frac{n}{2}$PE}} &\makecell{\vspace{-0.1cm}\\ $\alpha \leq 0.00066$, $\mathcal{M}^{A}$} & \multirow{2}{*}{}& \hspace{0.07cm}\makecell{\vspace{-0.2cm}\\ \includegraphics[angle=0,width=0.2\linewidth]{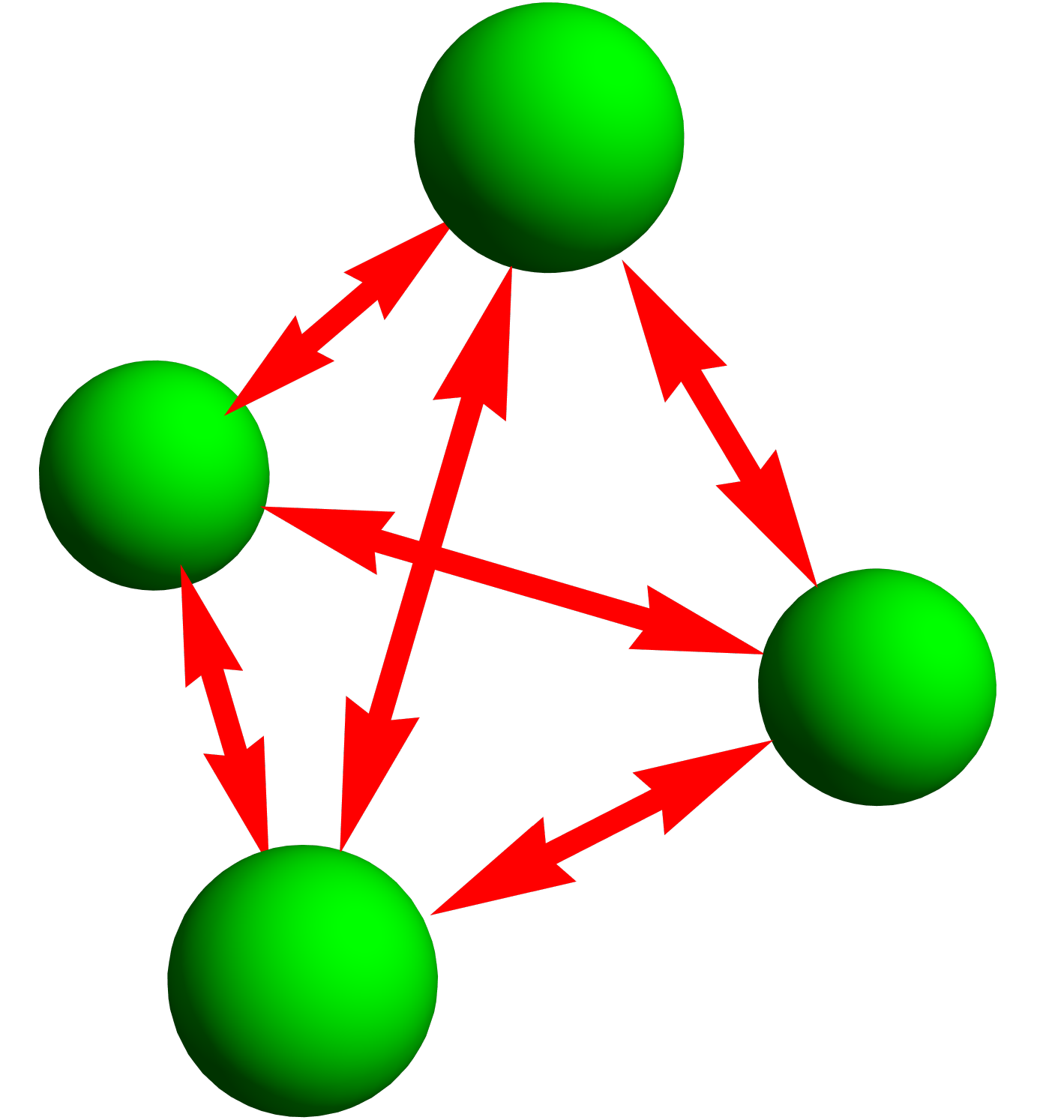}}\\
 &
 \begin{tikzpicture}
 \draw [dashed] (0,1) -- (2.7,1);
\end{tikzpicture} 
 &
 &
 \begin{tikzpicture}
  \draw [dashed] (0,1) -- (1.8,1);
\end{tikzpicture} \\
 & \makecell{$\alpha\geq 0.00154$, $\mathcal{M}^{A}$\\ or $\forall \alpha$, $\mathcal{M}^{E}$ \\ \vspace{-0.1cm}} & 
 & \hspace{0.1cm} \includegraphics[angle=0,width=0.2\linewidth]{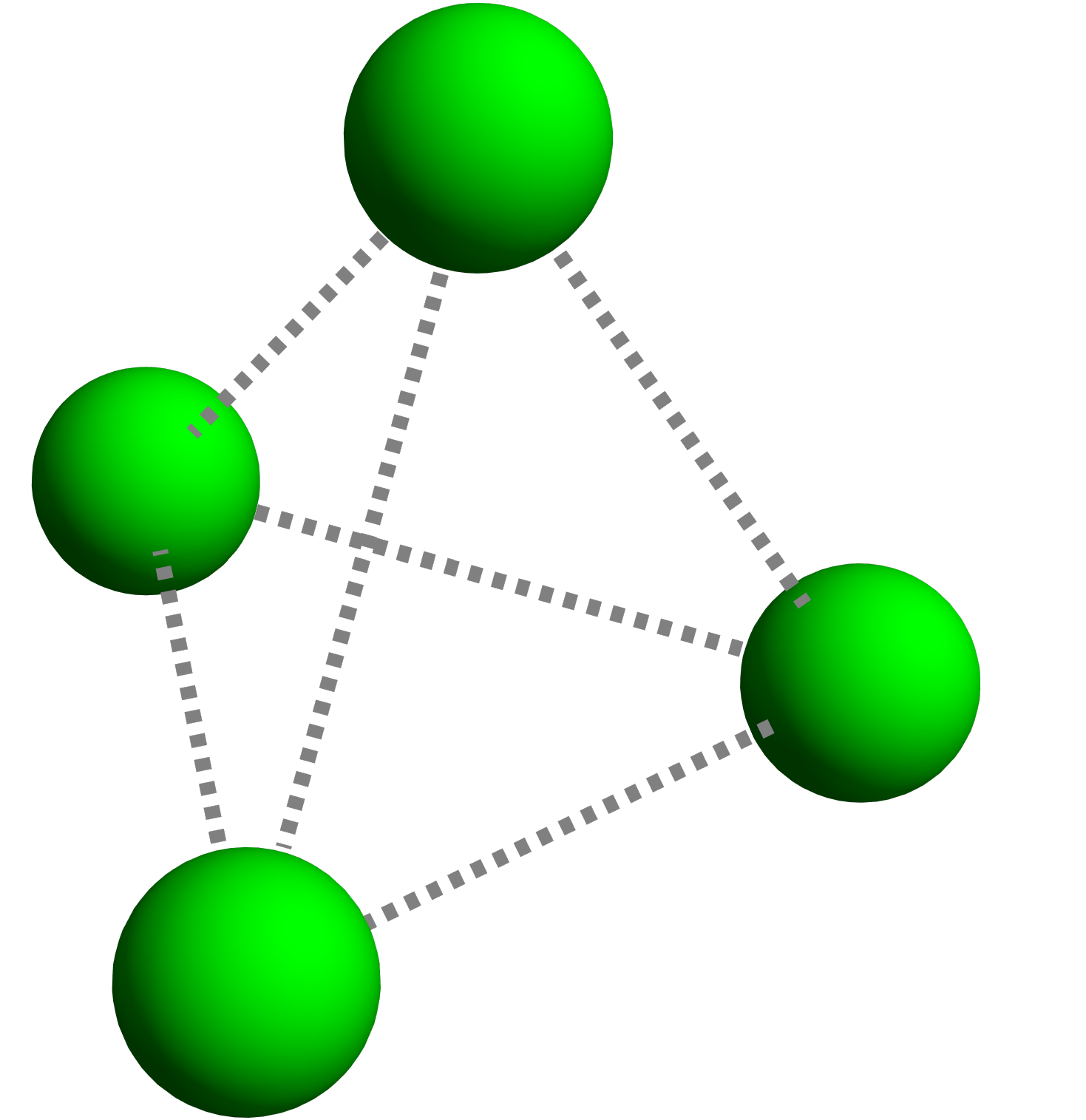}\vspace{0.2cm} \\
\hline
\multirow{7}{*}{\makecell{RPE}} 
&\makecell{\vspace{-0.1cm}\\ $\alpha\leq 0.05857$, $\mathcal{M}^{A}$} 
&\makecell{\vspace{-0.2cm}\\ \includegraphics[angle=0,width=0.2\linewidth]{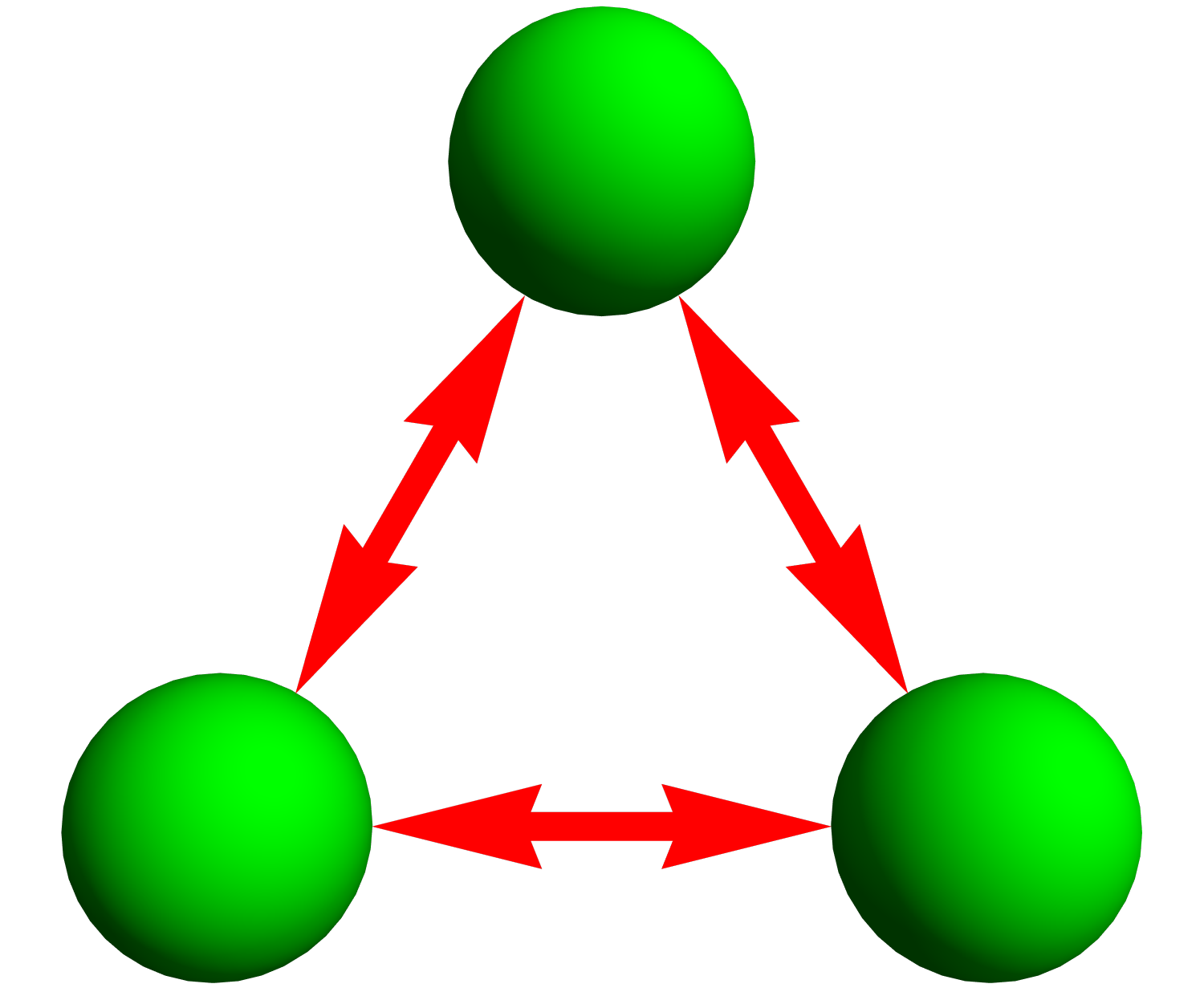}} 
&\multirow{7}{*}{ \hspace{0.07cm}\includegraphics[angle=0,width=0.2\linewidth]{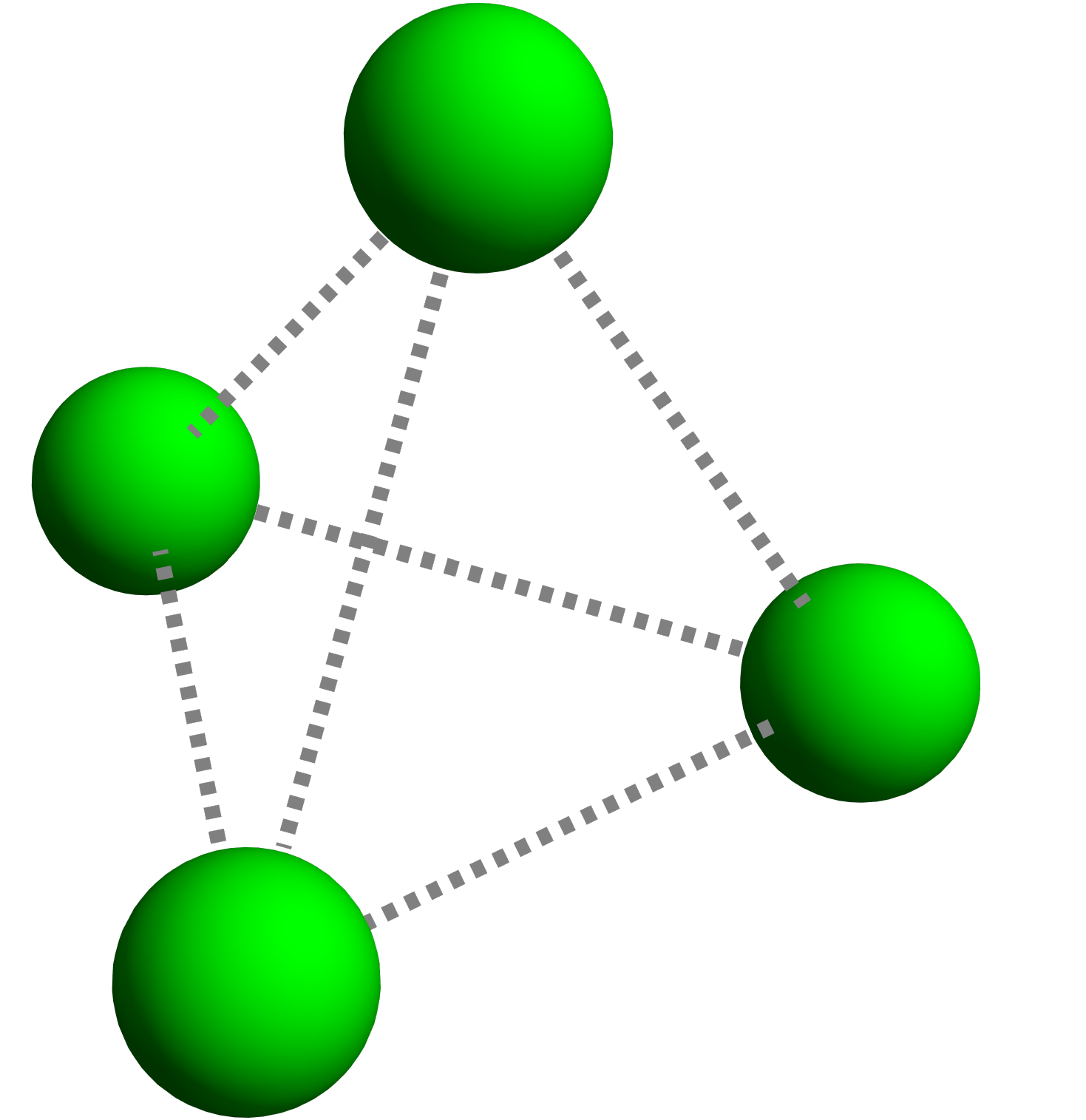}}\\
 &
 \begin{tikzpicture}
 \draw [dashed] (0,1) -- (2.7,1);
\end{tikzpicture} 
 &
 \begin{tikzpicture}
  \draw [dashed] (0,1) -- (1.8,1);
\end{tikzpicture} 
 &\\
 & \makecell{ $\alpha\geq 0.06209$, $\mathcal{M}^{A}$ \\ or $\forall \alpha$,  $\mathcal{M}^{E}$\\ \vspace{-0.1cm} }
 & \includegraphics[angle=0,width=0.2\linewidth]{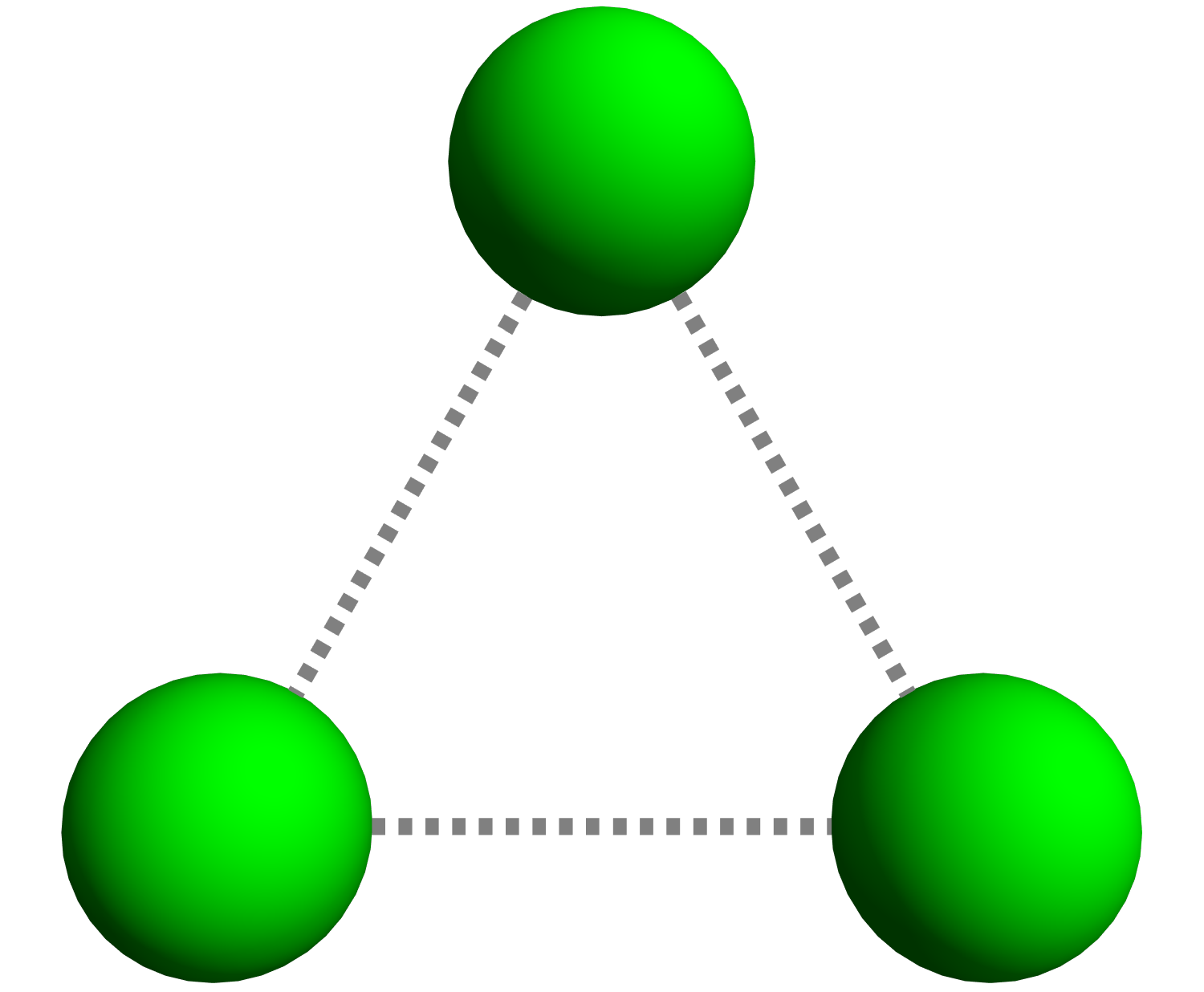}\vspace{0.2cm}
 &\\
\hline
\multirow{20}{*}{\makecell{SRPE}} & \makecell{\vspace{-0.2cm}\\ $\forall \alpha$,  $\mathcal{M}^{A}$} 
&\multirow{8}{*}{\includegraphics[angle=0,width=0.2\linewidth]{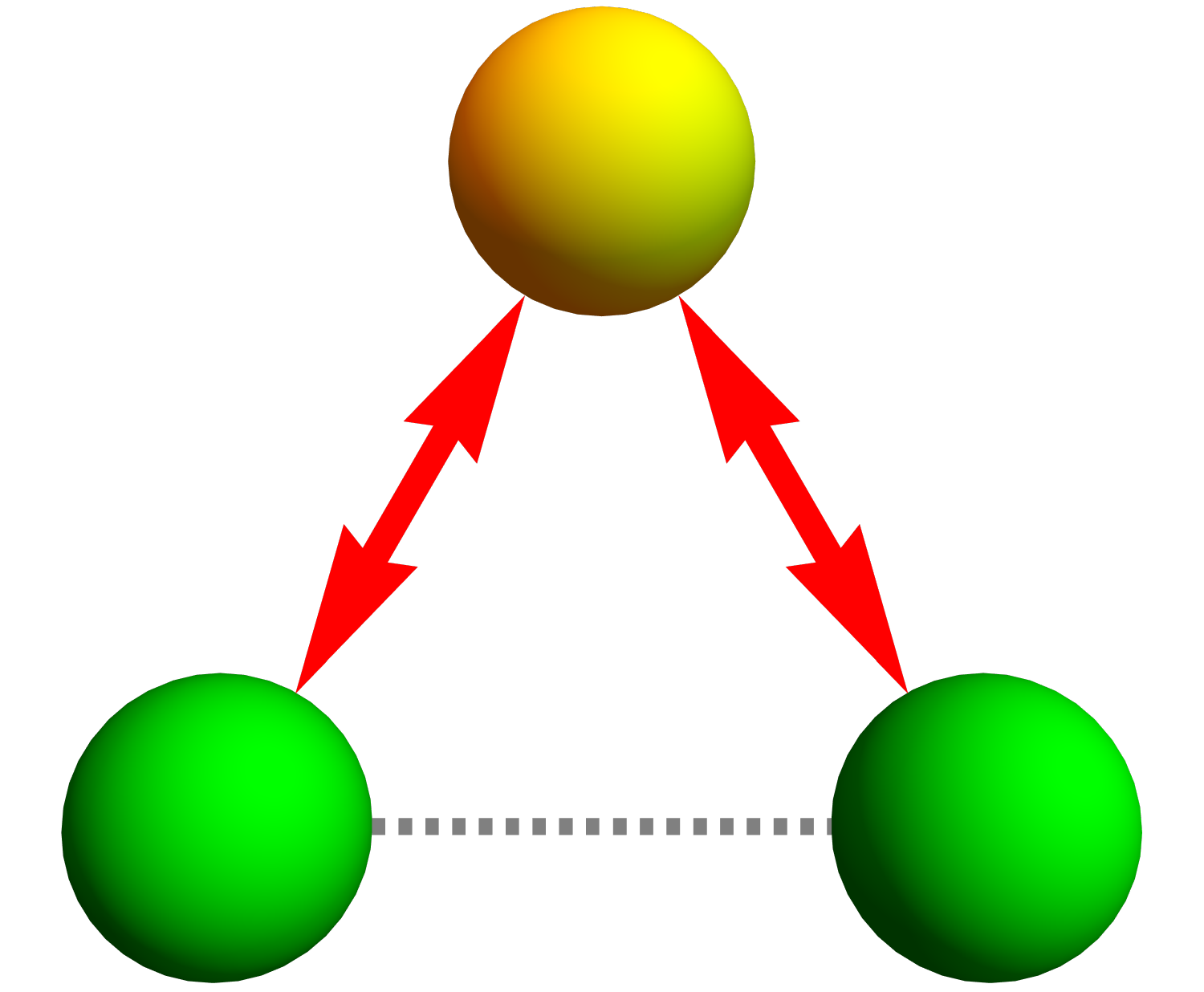}} &
\hspace{0.07cm}
\makecell{\vspace{-0.2cm}\\ \includegraphics[angle=0,width=0.2\linewidth]{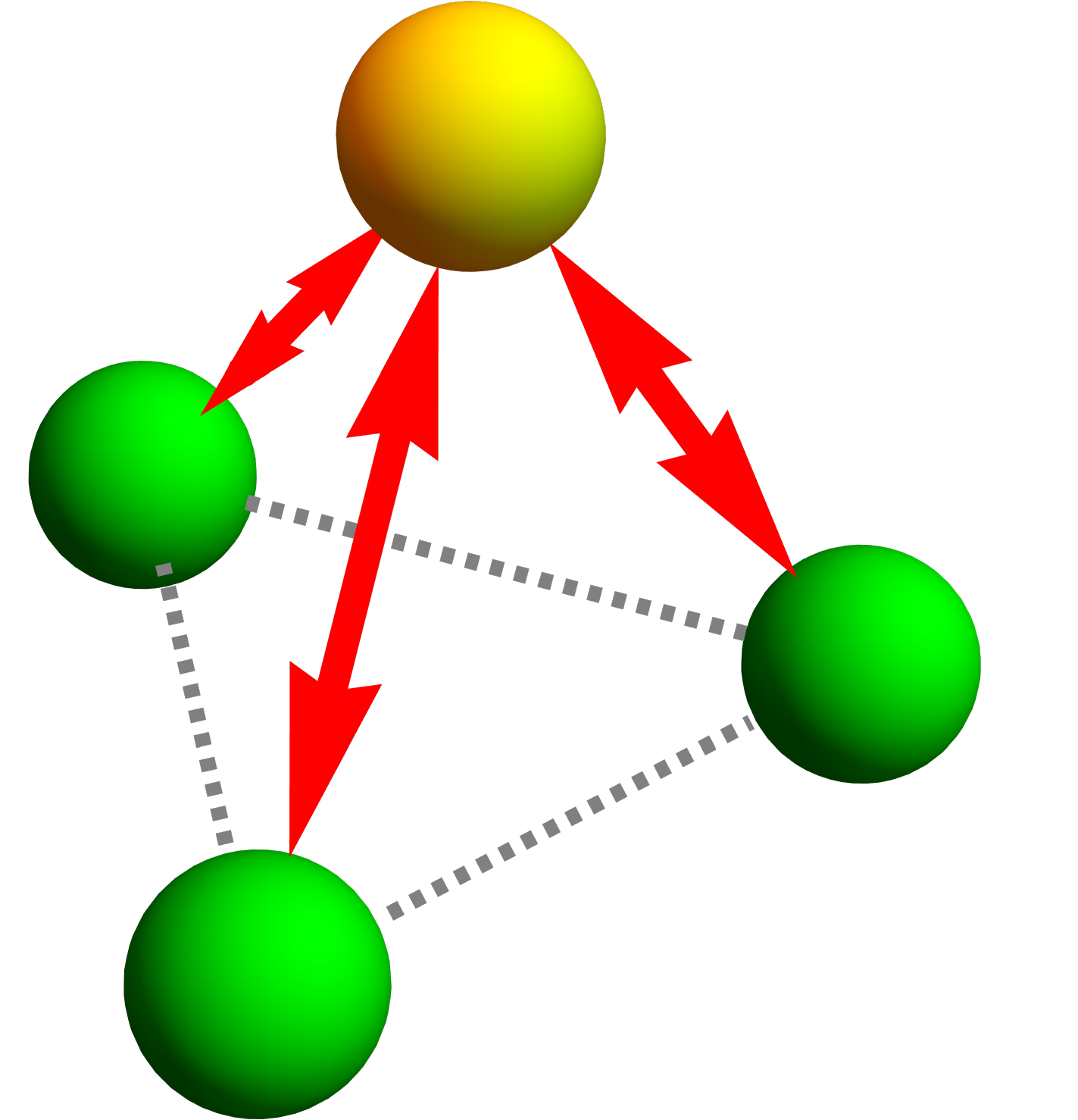}}\\
&
 \begin{tikzpicture}
 \draw [dashed] (0,1) -- (2.7,1);
\end{tikzpicture} 
 &

 &
 \begin{tikzpicture}
  \draw [dashed] (0,1) -- (1.8,1);
\end{tikzpicture} \\
 & \makecell{\vspace{-0.01cm}\\ $\forall \alpha$,  $\mathcal{M}^{E}$\\ \vspace{-0.01cm}}
& &
\multirow{6}{*}{\hspace{0.07cm}\includegraphics[angle=0,width=0.2\linewidth]{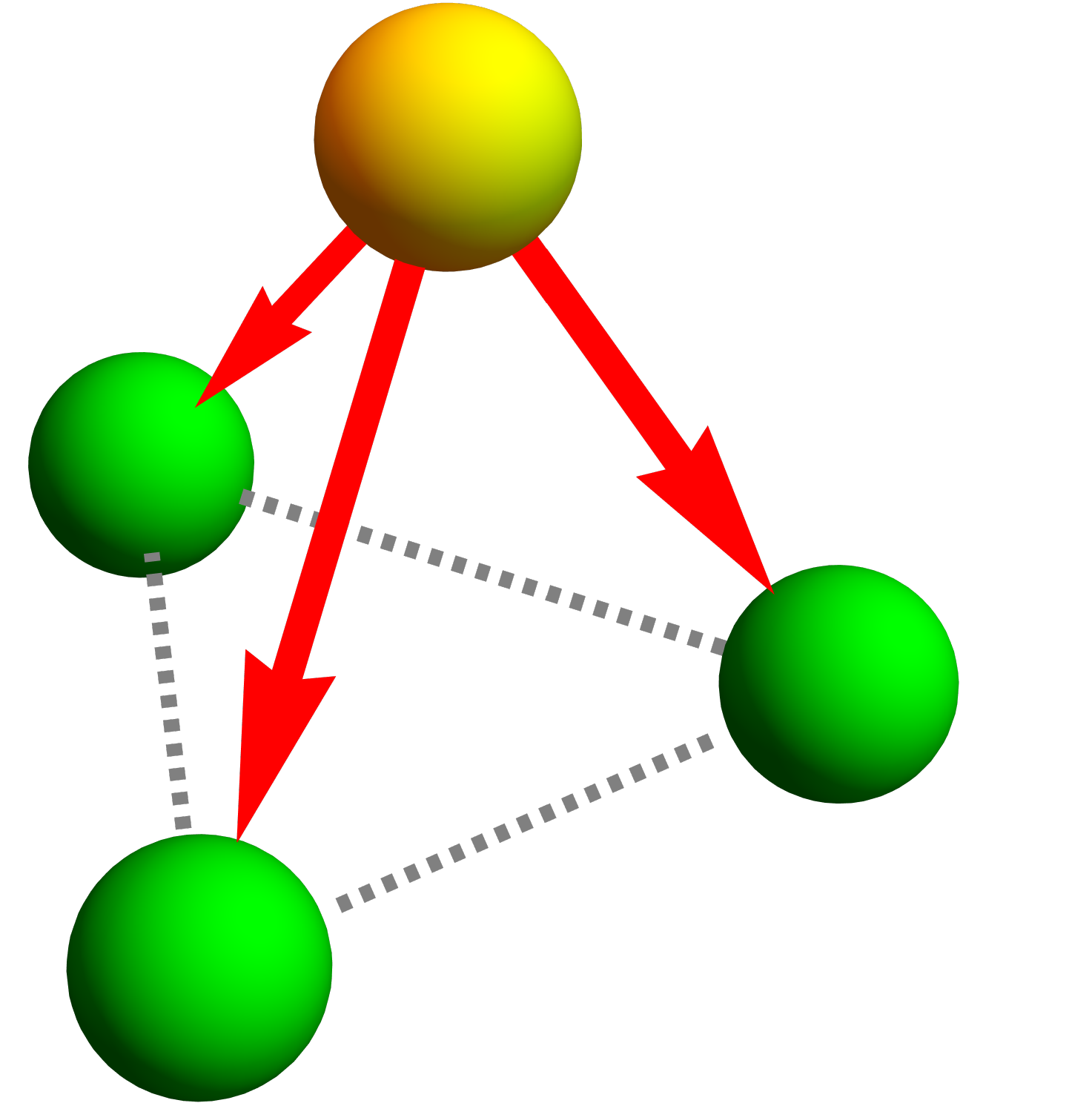}}\\
&
 \begin{tikzpicture}
 \draw [dashed] (0,1) -- (2.7,1);
\end{tikzpicture} 
 &
 \begin{tikzpicture}
  \draw [dashed] (0,1) -- (1.8,1);
\end{tikzpicture} 
 &
\\
&  \makecell{\vspace{-0.2cm}\\$\alpha<\frac{1}{2}$, $\mathcal{M}_3$\\ or $\alpha<\frac{1}{3}$, $\mathcal{M}_2$\\\vspace{-0.2cm}}
&\multirow{5}{*}{\includegraphics[angle=0,width=0.2\linewidth]{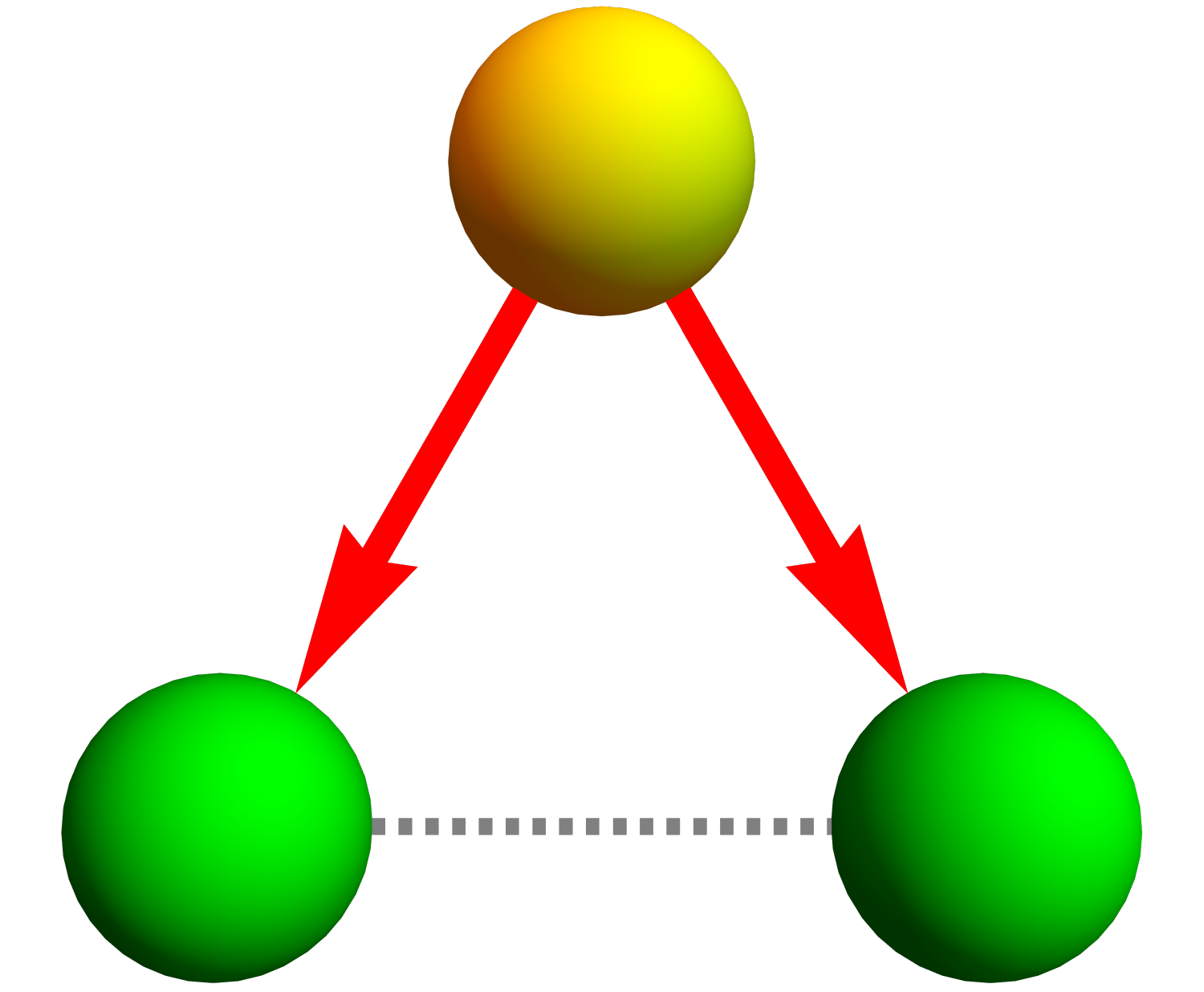}} & \\
&
 \begin{tikzpicture}
 \draw [dashed] (0,1) -- (2.7,1);
\end{tikzpicture} 
 &
 &
 \begin{tikzpicture}
  \draw [dashed] (0,1) -- (1.8,1);
\end{tikzpicture} \\
&\makecell{\vspace{-0.2cm}\\ $\alpha=\frac{1}{2}$, $\mathcal{M}_3$\\ or $\frac{1}{3}\leq \alpha <\frac{1}{2}$, $\mathcal{M}_2$\\ \vspace{-0.2cm}}
& & \multirow{5}{*}{\hspace{0.07cm}\includegraphics[angle=0,width=0.2\linewidth]{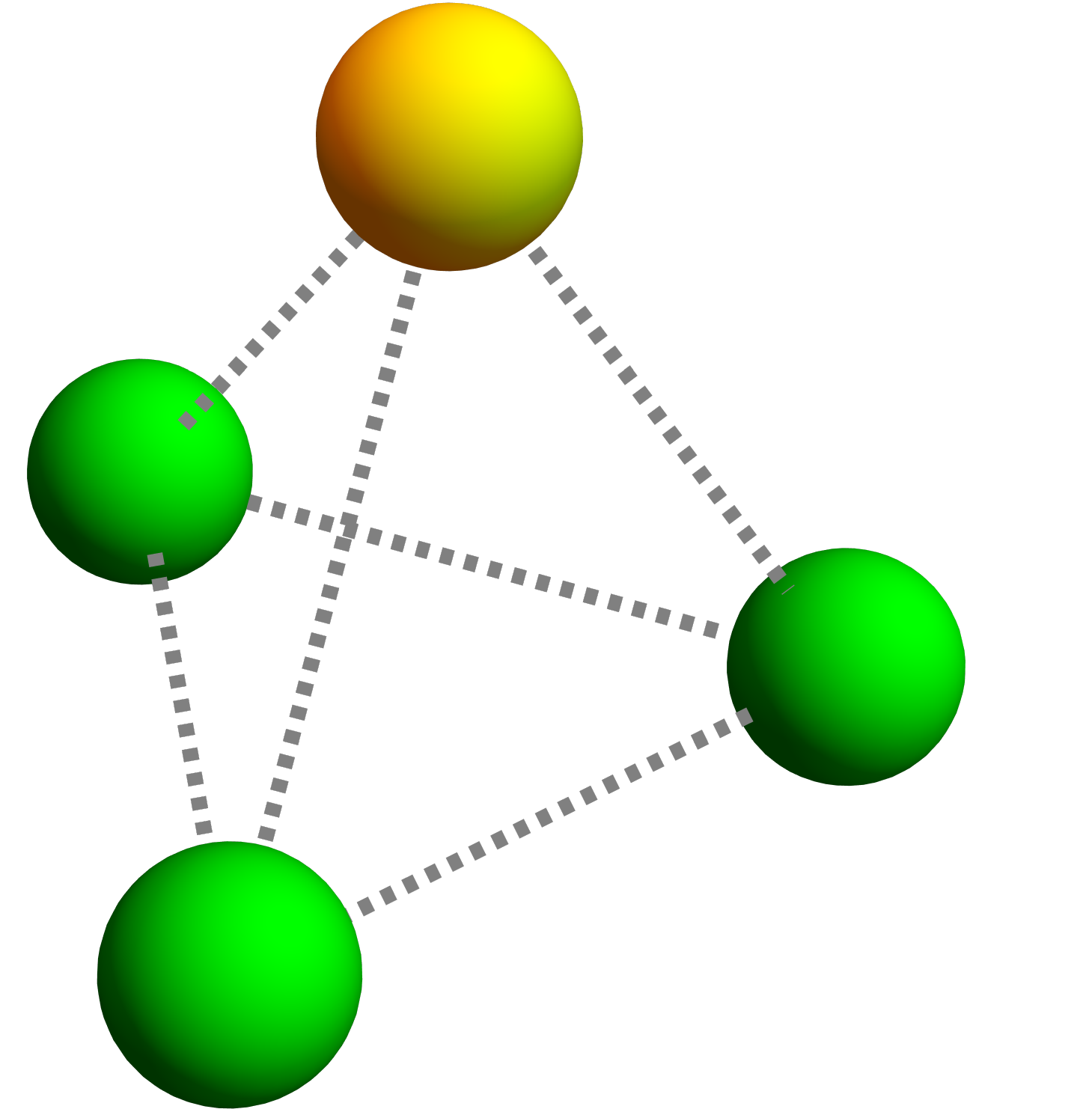}}\\
&
 \begin{tikzpicture}
 \draw [dashed] (0,1) -- (2.7,1);
\end{tikzpicture} 
 &
  \begin{tikzpicture}
  \draw [dashed] (0,1) -- (1.8,1);
\end{tikzpicture}
 &
 \\
 &
 \makecell{$\alpha=\frac{1}{2}$, $\mathcal{M}_2$\\ \vspace{-0.2cm}} &
 \makecell{\includegraphics[angle=0,width=0.2\linewidth]{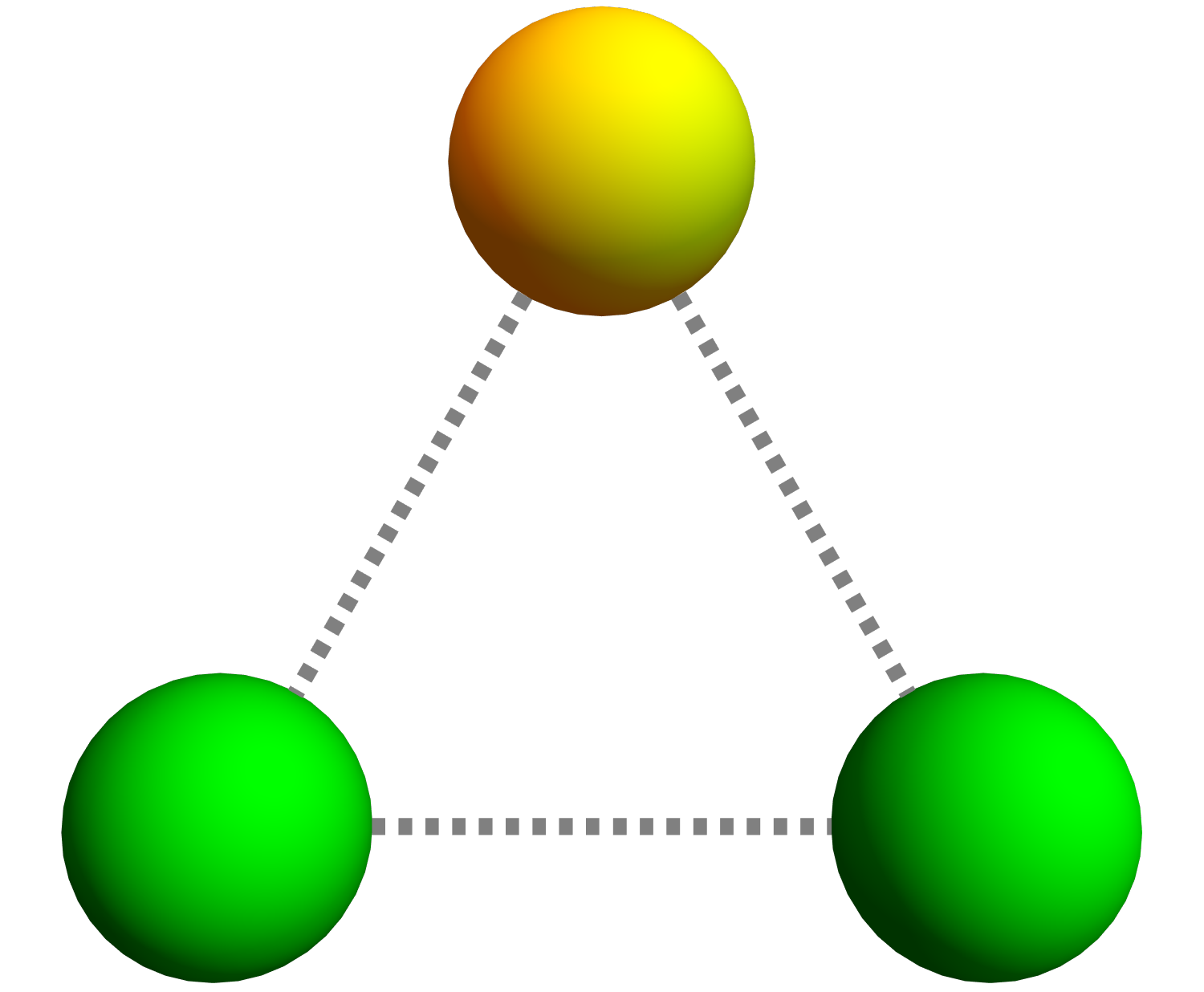}\\ \vspace{-0.2cm}} & \\
\hline
\end{tabular}
\caption{The classes of three small networks. 
All three states considered have $\alpha\in(0,1/2]$. One-way arrows denote one party steering the other, while double arrows indicate mutual steering between two parties. The R$\frac{n}{2}$PE and RPE states have numerically unclassified regions for $0.00066<\alpha<0.00154$ and $0.05857<\alpha<0.06209$, respectively.
}\label{table-figure}
\end{figure}

In this section, we study the properties of small networks ($n=3,4$) in the three scenarios, R$\frac{n}{2}$PE, RPE, and SRPE. We find that the monogamy and shareability of EPR-steering is dependent on the measurement schemes and the state parameter $\alpha$. We summarize the results for various small network configurations in Fig.~\ref{table-figure}.
There are four different cases with specific conditions as follows:
\begin{enumerate}
\item Each party can simultaneously steer the states of all other parties because there exists symmetry with respect to each parties. (R$\frac{n}{2}$PE and RPE states)
\item One party  (Alice) can simultaneously steer the states of all other parties (Bob1, Bob2, Bob3), and each Bob can steer Alice's state, but the Bobs cannot steer one-another. (SRPE state)
\item Alice can simultaneously steer all the Bobs' states, but no Bob can steer Alice's state, and the Bobs cannot steer one-another. (SRPE state)
\item No party can steer any other party. (R$\frac{n}{2}$PE, RPE and SRPE states)
\end{enumerate}

\begin{figure}[t]\centering
\includegraphics[angle=0,width=0.87\linewidth]{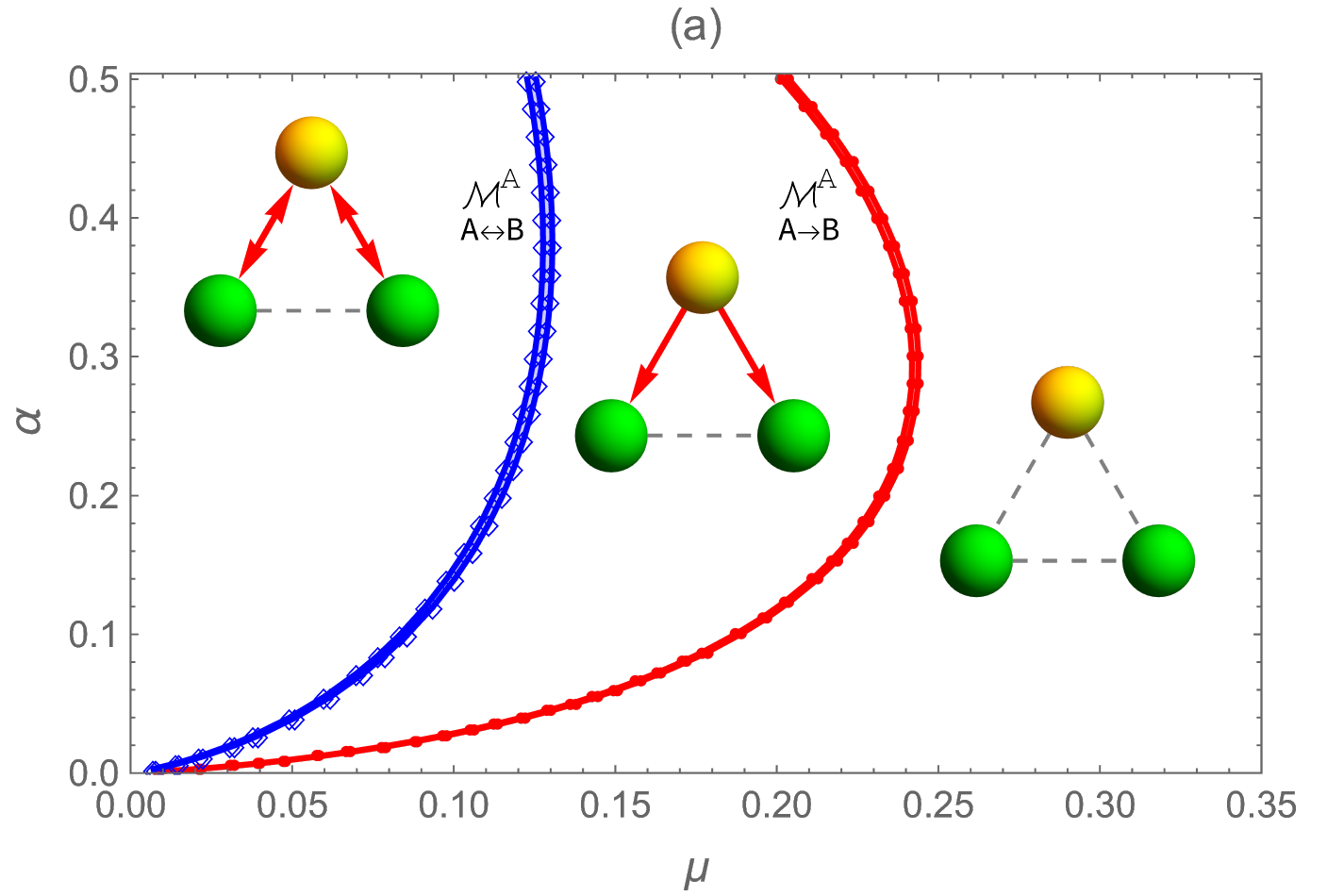}
\hspace{1.5cm}
\includegraphics[angle=0,width=0.87\linewidth]{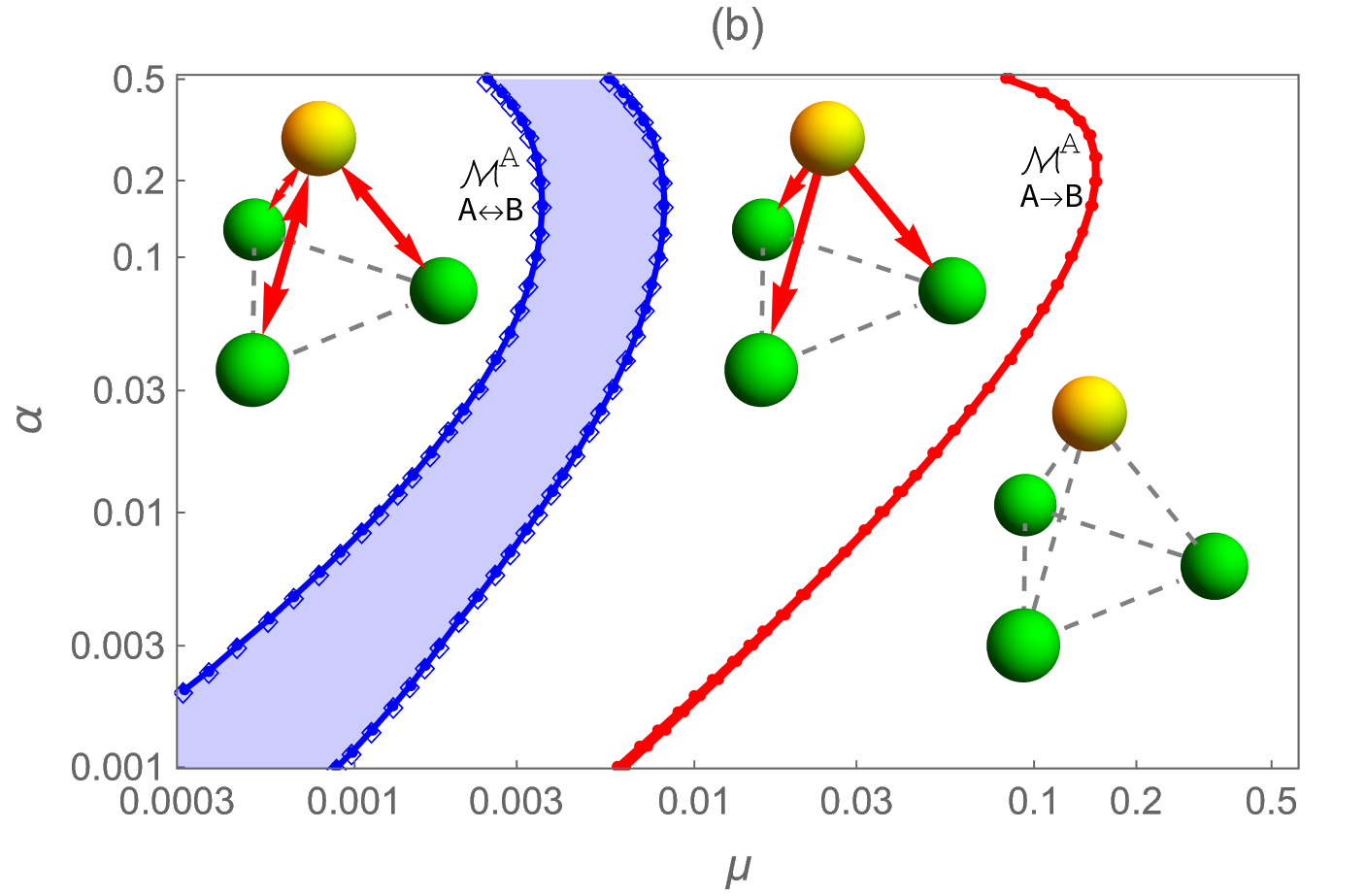}
\caption{Small steering networks of SRPE state with noise $\mu$. 
The measurement scheme $\mathcal{M}^A$ (all projective measurements) is considered. The steerable region is the left-hand side of each curve.  Subfigure (a) shows network properties of the tripartite SRPE state with a linear scale on both axes, while subfigure (b) displaying the properties of the 4-partite SRPE state  with a log scale on both axes.
}\label{fig-srpe-noise-network}
\end{figure}

\begin{figure}[h]
\begin{tabular}{|M|M|M|}
\hline
{Scenario} & {Condition}  & $n> 4$ \\
\hline
\multirow{6}{*}{\makecell{SRPE}} 
&\makecell{\vspace{-0.1cm}\\ $\alpha<\frac{1}{n-1}$, $\mathcal{M}_2$} 
&\multirow{6}{*}{ \hspace{0.07cm}\includegraphics[angle=0,width=0.35\linewidth]{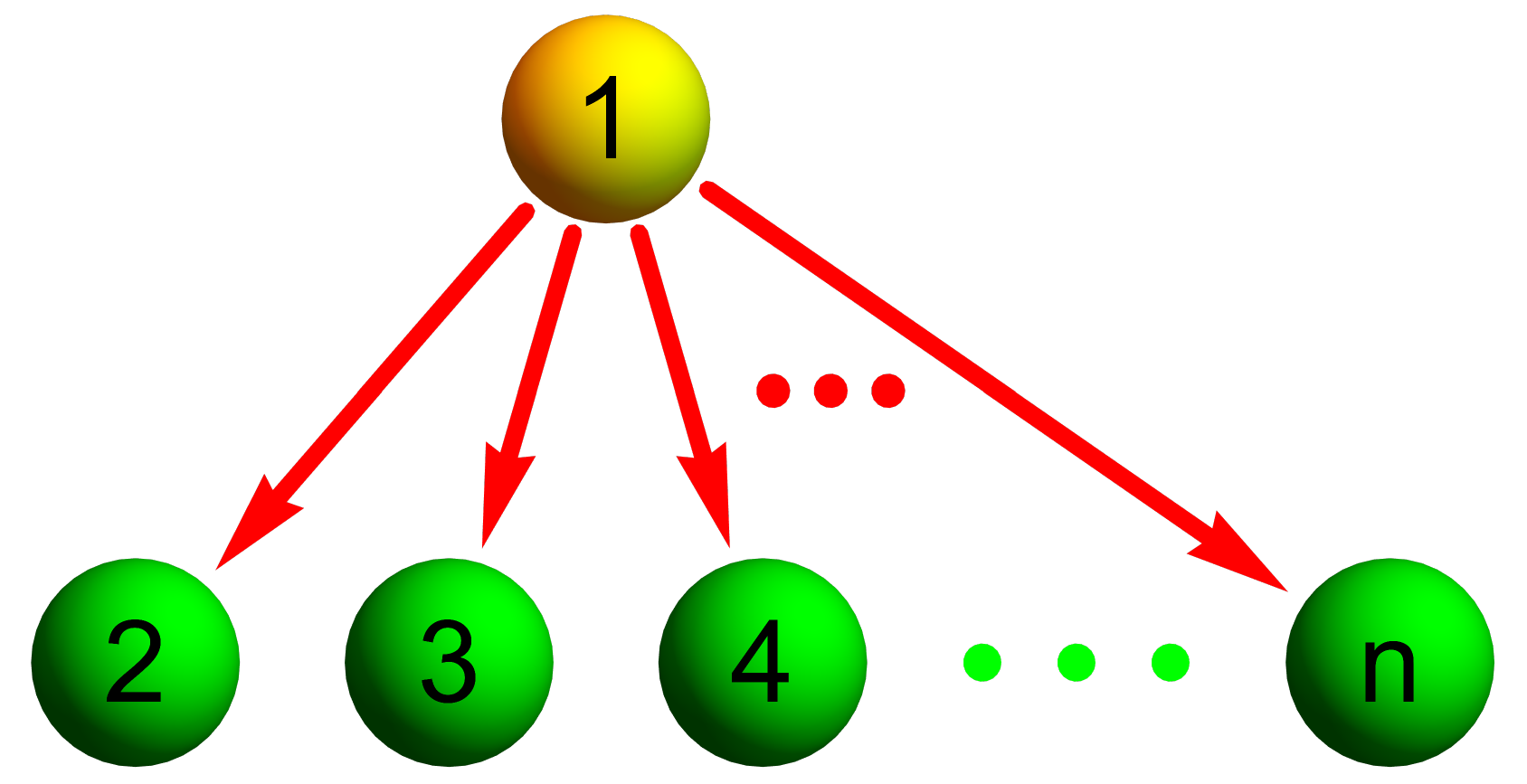}}\\
 &
 \begin{tikzpicture}
 \draw [dashed] (0,1) -- (2.7,1);
\end{tikzpicture} 
 &\\
 & \makecell{ $\alpha<\frac{2}{n}$, $\mathcal{M}_3$ }
 &\\
 &
 \begin{tikzpicture}
 \draw [dashed] (0,1) -- (2.7,1);
\end{tikzpicture} 
 &\\
 & \makecell{ $\alpha<\frac{\pi ^2}{4(n-2)+\pi ^2}$,  $\mathcal{M}^{E}$\\ \vspace{-0.1cm} }
 &\\
\hline
\end{tabular}
\caption{Steering properties of the SRPE state with $n>4$. If the stated condition is not satisfied then no steering is possible.
}\label{table-figure-2}
\end{figure}

It is noteworthy that in Ref~\cite{Paul20}, the authors demonstrated that in a three-party system ($n = 3$), it is impossible for all three pairs of qubits to violate the three-setting CJWR linear steering inequality~\cite{Caval09}. 
However, in our RPE scenario, when $n=3$, $0<\alpha\leq 0.05857$ and each party can perform all projective measurements $\mathcal{M}^{A}$, then there exists bipartite steering between all three pair parties which can steer each other as shown in Fig.~\ref{table-figure}. 
For $n=4$, a similar situation occurs in R$\frac{n}{2}$PE scenario. (Recall that $n=3$ does not apply to the R$\frac{n}{2}$PE scenario.) This reveals that more shareability of steering can be shown by increasing the number of measurement settings.

In contrast to R$\frac{n}{2}$PE and RPE states, the 
SRPE state exhibits stronger steering properties, making it more likely to preserve steering under high levels of noise. Consequently, we select SRPE state as the focus of our quantitative analysis of steering networks in noisy environments.
We study steering properties of small ($n=3,4$) networks of SRPE state with noise $\mu$ under all projective measurements $\mathcal{M}^A$. 
As shown in Fig.~\ref{fig-srpe-noise-network}, despite the presence of noise, the  SRPE state maintains its three steering structures for $n=3,4$, similar to when there is no noise present. 
Especially in the case of $n=3$, these results show that the SRPE state is indeed very robust against noise. 
As the noise $\mu$ increases, the maximum number of Bobs that Alice can simultaneously steer decreases.

Regarding the steering properties of networks with $n>4$, as depicted in Figs.~\ref{steering-half}~and~\ref{steering-rpe}, it is observed that when $\alpha\geq 0.0001$, there is no steering in either the R$\frac{n}{2}$PE or RPE scenarios. It is safe to assume that this is also true when $\alpha<0.0001$. 
For SRPE scenario with $n>4$, however, Alice is able to steer an arbitrary number of Bobs simultaneously, provided that the relevant conditions on $\alpha$ are fulfilled,  as illustrated in Fig.~\ref{table-figure-2}.
As more measurements are performed, the constraint on parameter $\alpha$ becomes less stringent. It would be even less stringent if Alice could perform all projective measurements, $\mathcal{M}^A$, but the bound is not known analytically.

\section{Conclusion and Discussion}
\label{sec:Conclusion}

In this paper, the steerability of $n$-party 2-producible entangled states was investigated in three scenarios: R$\frac{n}{2}$PE, RPE, and SRPE. We obtained numerous results analytically, by first deriving the necessary and sufficient EPR-steering criteria for two-qubit X-states with the restriction $|t_x|=|t_y|$ for a family of measurement schemes. A particularly remarkable result is for SRPE (Semi-Random Pair Entanglement), which is produced from a single two-qubit state, with ancillary separable states. For this SRPE state, one party, who holds one half of the entangled pair, can steer any one of the other $n-1$ parties, with no upper limit on $n$. Finally, we studied the steering properties in the three scenarios in small networks, revealing various steering configurations under different conditions. It was observed that the SRPE state exhibits remarkable resilience to the effects of noise.

Our investigation only utilized one type of two-qubit entangled state in the three scenarios. However, it is worth noting that the steerabilities of various $n$-party 2-producible entangled states could be explored by employing different types of two-qubit entangled states at the same time, and also higher-dimensional entangled states. These states may demonstrate unique steering properties and shareability. Additionally, $n$-party $k$-producible entangled states $(k\geq3)$ are expected to exhibit stronger steerability.

We also presented numerical results (upper and lower bounds) for the case of the measurement scheme $\mathcal{M}^{A}$ 
(all projective measurements). These results have some interesting limitations. For instance, the steering upper bound of the bipartite reduced SRPE state $\varrho_{\rm SRPE}^{2\mathrm{r}}$ cannot be obtained using the numerical approach from Ref.~\cite{Ngu19}. Even in the presence of noise in the SRPE state, the difference between the upper and lower bounds becomes significant, as shown in Figs.~\ref{steering-srpe}~and~\ref{fig-srpe-noise-network}, and computing the bounds is time-consuming. Detecting steering efficiently thus remains a difficult task in general, and finding a necessary and sufficient analytical criterion for all two-qubit states is still an open problem.
It is also interesting to note that in all three scenarios without noise, states constructed from smaller values of $\alpha$ lead to larger steerable region from Alice to Bob. 
Intuitively, this is because Bob’s reduced state increases in purity with decreasing $\alpha$. 
This is because probabilistically swapping in the unentangled pure state that becomes closer to Bob’s reduced state in this limit, sufficiently preserves the entanglement present in the two-party reduced state.
Moreover, making Bob’s reduced state closer to a pure state is equivalent to increasing the probability of the pure steered states, in the noiseless case, which we have recently found to be particularly relevant to maintaining steerability, even when the amount of entanglement present is reduced~\cite{song22}.
Another interesting avenue for future work would analyze whether this behaviour for small $\alpha$ values has implications for monogamy relations for EPR-steering~\cite{Reid13}, and whether such constraints (or their non-existence) can be shown in these limits. 

\section*{acknowledgement}

We thank an anonymous referee for drawing our attention to the connection of the setup in Section~\ref{sec:Rn2PE} with perfect matchings in graph theory.
Qiu-Cheng Song acknowledges support by a cotutelle Scholarship from Griffith University and University of Chinese Academy of Sciences. 
This work was supported by the ARC Centre of Excellence for 
Quantum Computation and Communication Technology (CQC2T), project number CE170100012. This research was supported by the Griffith University Gowonda HPC Cluster.

\section*{APPENDIX: The proof of Theorem 1}
\label{sec:appendixa}

This appendix contains the proof of Theorem~\ref{Theorem} using a technique employed in Ref.~\cite{Jon07}.

Let's consider X-states under the constraint $|t_x|=|t_y| =: t_\perp$. Alice chooses the measurement scheme
\begin{align}
\mathcal{M}_m^D=\{\sigma_z, \sigma_\theta\}_\theta,
\end{align}
where $\sigma_\theta$ is defined as
\begin{align}
\sigma_\theta=\cos{(\theta)}\sigma_x+\sin{(\theta)}\sigma_y,
\end{align}
with $\theta=\frac{l\pi}{m}$ and $l=0,1,2,\cdots,2m-1$,  and the ensemble for the measurement scheme is denoted by

\begin{align}
\mathcal{E}=\{\ketbra{\psi_l}d\mu(\psi_l)\},
\end{align} 
where  
\begin{align}
|\psi_l\rangle\langle\psi_{l}|
=&\frac{1}{2}\bigg[\openone+\sqrt{1-z^2}\sin\left[(2l+1)\frac{\pi }{2 m}\right]\sigma_x\notag\\
&+ \sqrt{1-z^2}\cos\left[(2l+1)\frac{\pi}{2 m}\right]\sigma_y+z\sigma_z\bigg]
\end{align}
and 
\begin{align}
d\mu(\psi_l)=\frac{1}{2m}{p}(z)dz.
\end{align}
It is important to note that finding the optimal probability distribution $p(z)$ is necessary to ensure that the ensemble $\mathcal{E}$ is optimal.

Let's first consider Alice to measure $\sigma_z$, which is described by the effect $E_{r\mid z}=\frac{1}{2}\left[\openone+(-1)^{r} \sigma_{z}\right]$, then Bob's states become
\begin{align}\label{x1}
\sigma_{\pm|z}
&=\frac{p_{\pm|z}}{2}\left[\openone+z_{\pm}\sigma_z\right],
\end{align}
with probabilities
\begin{align}\label{pp}
&p_{\pm|z}=\frac{1\pm a}{2},
\end{align}
where
\begin{align}\label{zz}
&z_\pm=\frac{b\pm t_z}{1\pm a}.
\end{align}
To reproduce these steered states optimally, Alice should divide $p(z)$ into two positive distributions
\begin{align}
p(z)=p_+(z)+p_-(z),
\end{align}
where the ensemble $\mathcal{E}$ with distributions $p_+(z)$ and $p_-(z)$ simulates conditioned states $\sigma_{+|z}$ and $\sigma_{-|z}$, respectively. With this strategy, Bob can obtain, on average, the following states:
\begin{align}\label{x2}
\sigma_{\pm|z}=&\sum_{l=0}^{2m-1}\int_{-1}^{+1}\ketbra{\psi_l}\frac{1}{2 m}{p}(z)dz\notag\\
=&\frac{1}{2}\left[\openone \int_{-1}^{+1}dzp_{\pm}(z)+\sigma_z \int_{-1}^{+1}dzp_{\pm}(z)z\right].
\end{align}
If the steered states can be simulated using the ensemble $\mathcal{E}$ and this strategy, we can obtain constraints on $p(z)$ by comparing Eqs.~\eqref{x1}~and~\eqref{x2}
\begin{align}
&\int_{-1}^{+1}dz p_{+}(z)=\frac{1+a}{2},\label{xa}\\
&\int_{-1}^{+1}dz p_{-}(z)=\frac{1-a}{2},\label{xb}\\
&\int_{-1}^{+1}dz p_{+}(z)z=\frac{b+t_z}{2},\label{xc}\\
&\int_{-1}^{+1}dz p_{-}(z)z=\frac{b-t_z}{2}.\label{xd}
\end{align}
If Alice were to measure $\sigma_\theta$ described by the effect $E_{r\mid \theta}=\frac{1}{2}\left[\openone+(-1)^{r}\cos{(\theta)}\sigma_{x}+(-1)^{r}\sin{(\theta)}\sigma_{x}\right]$, then Bob's conditioned states are given by:
\begin{align}\label{x3}
\sigma_{\pm|\theta }=\frac{1}{4}\left[\openone \pm t_x\cos{(\theta)}\sigma_x \pm t_y \sin{(\theta)}\sigma_y+b_z\sigma_z\right].
\end{align}
The optimal ensemble $\mathcal{E}^{\star}$ exhibits symmetry under rotations about the $z$-axis. Same as in Ref.~\cite{Jon07}, Alice's response function is described by the distribution
\begin{equation}\label{x4}
p(\pm | \sigma_\theta,(\beta,z))=\left\{\begin{aligned}
1,\quad  \mathrm{if}\quad \beta\in\left[\theta \mp \frac{\pi}{2}, \theta \pm \frac{\pi}{2}\right),\\
0,\quad  \mathrm{if}\quad \beta\in\left[\theta \pm \frac{\pi}{2}, \theta \mp \frac{\pi}{2}\right),
\end{aligned}\right.
\end{equation}
where $\beta=(2l+1)\pi /(2 m)$. 
Without loss of generality, we can set $\theta=0$, which simplifies Eq.~\eqref{x3} to
\begin{align}\label{x5}
\sigma_{\pm|x}=\frac{1}{4}\left[\openone \pm t_x\sigma_x+b_z\sigma_z\right].
\end{align}
By using the optimal ensemble $\mathcal{E}^\star$ with the strategy given in Eq.~\eqref{x4}, Bob's average states can be expressed as
\begin{align}\label{x6}
&\sum_{l=0}^{m-1}\int_{-1}^{+1}\ketbra{\psi_l}\frac{1}{2 m}{p}(z)dz\notag\\
=&\frac{1}{4}\left[\openone+M\int_{-1}^{1}\sqrt{1-z^2}p(z)dz\sigma_x+\int_{-1}^{1}p(z)z dz\sigma_z\right],
\end{align}
where 
\begin{align}
M=\frac{1}{m}\csc \left(\frac{\pi }{2 m}\right).
\end{align}
This expression is equal to Eq.~(4.15) in Ref.~\cite{Jones11}. We need to find the optimal $p(z)$ that maximizes the expression $\int_{-1}^{1}\sqrt{1-z^2}p(z)dz$. Conditioned on the constraints given in Eqs.~(\ref{xa})-(\ref{xd}) and by using Lagrange multiplier techniques, the optimal $p(z)$ is
\begin{align}
p^\star(z)=p_{+|z}\delta\left(z-z_+\right)+p_{-|z}\delta\left(z-z_-\right),
\end{align}
where $\delta\left(z-z_\pm\right)$ is the Dirac delta function, $p_{\pm|z}$ and $z_{\pm}$ are defined in Eqs.~\eqref{pp}~and~\eqref{zz}, respectively. Using the optimal $p^\star(z)$, the expression in Eq.~\eqref{x6} becomes
\begin{align}\label{x7}
&\frac{1}{4}\left[\openone \pm M\left(p_{+|z}\sqrt{1-z_+^2}+p_{-|z}\sqrt{1-z_-^2}\right)\sigma_x+b_z\sigma_z\right].
\end{align}
Comparing Eqs.~\eqref{x5}~and~\eqref{x7}, it can be observed that Alice's optimal strategy fails to reproduce $\sigma_{\pm|x}$ if and only if the inequality in Eq.~\eqref{xsteeringineqnnn} holds.

\input{The_shareability_of_steering_in_two-producible_states.bbl}
\bibliographystyle{apsrev4-1}

\end{document}

%% file: The_shareability_of_steering_in_two-producible_states.bbl
%